\documentclass[prl,aps,twocolumn,superscriptaddress,showpacs,nofootinbib]{revtex4-1}
\usepackage{natbib}
\usepackage{amscd}
\usepackage{amsmath,amssymb,amsthm,mathrsfs,amsfonts,dsfont}
\usepackage{subfigure, epsfig}
\usepackage{braket}
\usepackage{bm}
\usepackage{enumerate}

\usepackage{graphicx}
\usepackage{epsfig}
\usepackage{epstopdf}
\usepackage{subfigure}
\usepackage{color}
\usepackage[10pt]{moresize}

\newcommand\oprod[2]{\ensuremath{|#1\rangle\langle#2|}}
\newcommand\mean[1]{\ensuremath{\langle #1 \rangle}}

\usepackage{url}
\usepackage{array}
\newcommand{\PreserveBackslash}[1]{\let\temp=\\#1\let\\=\temp}
\newcolumntype{C}[1]{\rangle{\PreserveBackslash\centering}p{#1}}
\newcolumntype{R}[1]{\rangle{\PreserveBackslash\raggedleft}p{#1}}
\newcolumntype{L}[1]{\rangle{\PreserveBackslash\raggedright}p{#1}}

\begin{document}

\title{Sending-or-Not-Sending with Independent Lasers: Secure Twin-Field Quantum Key Distribution Over 509 km}

\author{Jiu-Peng Chen}
\author{Chi Zhang}
\affiliation{Shanghai Branch, National Laboratory for Physical Sciences at Microscale and Department of Modern Physics, University of Science and Technology of China, Shanghai 201315, P.~R.~China}
\affiliation{Shanghai Branch, CAS Center for Excellence and Synergetic Innovation Center in Quantum Information and Quantum Physics, University of Science and Technology of China, Shanghai 201315, P.~R.~China}

\author{Yang Liu}
\affiliation{Shanghai Branch, National Laboratory for Physical Sciences at Microscale and Department of Modern Physics, University of Science and Technology of China, Shanghai 201315, P.~R.~China}
\affiliation{Shanghai Branch, CAS Center for Excellence and Synergetic Innovation Center in Quantum Information and Quantum Physics, University of Science and Technology of China, Shanghai 201315, P.~R.~China}
\affiliation{Jinan Institute of Quantum Technology, Jinan, Shandong 250101, P.~R.~China}

\author{Cong Jiang}
\affiliation{State Key Laboratory of Low Dimensional Quantum Physics, Department of Physics, Tsinghua University, Beijing 100084, P.~R.~China}

\author{Weijun Zhang}
\affiliation{State Key Laboratory of Functional Materials for Informatics, Shanghai Institute of Microsystem and Information Technology, Chinese Academy of Sciences, Shanghai 200050, P.~R.~China}

\author{Xiao-Long Hu}
\affiliation{State Key Laboratory of Low Dimensional Quantum Physics, Department of Physics, Tsinghua University, Beijing 100084, P.~R.~China}

\author{Jian-Yu Guan}
\affiliation{Shanghai Branch, National Laboratory for Physical Sciences at Microscale and Department of Modern Physics, University of Science and Technology of China, Shanghai 201315, P.~R.~China}
\affiliation{Shanghai Branch, CAS Center for Excellence and Synergetic Innovation Center in Quantum Information and Quantum Physics, University of Science and Technology of China, Shanghai 201315, P.~R.~China}

\author{Zong-Wen Yu}
\affiliation{State Key Laboratory of Low Dimensional Quantum Physics, Department of Physics, Tsinghua University, Beijing 100084, P.~R.~China}
\affiliation{Data Communication Science and Technology Research Institute, Beijing 100191, P.~R.~China}

\author{Hai Xu}
\affiliation{State Key Laboratory of Low Dimensional Quantum Physics, Department of Physics, Tsinghua University, Beijing 100084, P.~R.~China}

\author{Jin Lin}
\affiliation{Shanghai Branch, National Laboratory for Physical Sciences at Microscale and Department of Modern Physics, University of Science and Technology of China, Shanghai 201315, P.~R.~China}
\affiliation{Shanghai Branch, CAS Center for Excellence and Synergetic Innovation Center in Quantum Information and Quantum Physics, University of Science and Technology of China, Shanghai 201315, P.~R.~China}

\author{Ming-Jun Li}
\author{Hao Chen}
\affiliation{Corning Incorporated, Corning, New York 14831, USA}

\author{Hao Li}
\author{Lixing You}
\author{Zhen Wang}
\affiliation{State Key Laboratory of Functional Materials for Informatics, Shanghai Institute of Microsystem and Information Technology, Chinese Academy of Sciences, Shanghai 200050, P.~R.~China}

\author{Xiang-Bin Wang}
\affiliation{Shanghai Branch, CAS Center for Excellence and Synergetic Innovation Center in Quantum Information and Quantum Physics, University of Science and Technology of China, Shanghai 201315, P.~R.~China}
\affiliation{Jinan Institute of Quantum Technology, Jinan, Shandong 250101, P.~R.~China}
\affiliation{State Key Laboratory of Low Dimensional Quantum Physics, Department of Physics, Tsinghua University, Beijing 100084, P.~R.~China}

\author{Qiang Zhang}
\author{Jian-Wei Pan}
\affiliation{Shanghai Branch, National Laboratory for Physical Sciences at Microscale and Department of Modern Physics, University of Science and Technology of China, Shanghai 201315, P.~R.~China}
\affiliation{Shanghai Branch, CAS Center for Excellence and Synergetic Innovation Center in Quantum Information and Quantum Physics, University of Science and Technology of China, Shanghai 201315, P.~R.~China}

\begin{abstract}
Twin field quantum key distribution promises high key rates at long distance to beat the rate distance limit. Here, applying the sending or not sending TF QKD protocol, we experimentally demonstrate a secure key distribution breaking the absolute key rate limit of repeaterless QKD over 509 km, 408 km ultra-low loss optical fibre and 350 km standard optical fibre. Two independent lasers are used as the source with remote frequency locking technique over 500 km fiber distance; Practical optical fibers are used as the optical path with appropriate noise filtering; And finite key effects are considered in the key rate analysis. The secure key rates obtained at different distances are more than 5 times higher than the conditional limit of repeaterless QKD, a bound value assuming the same detection loss in the comparison. The achieved secure key rate is also higher than that a traditional QKD protocol running with a perfect repeaterless QKD device and even if an infinite number of sent pulses. Our result shows that the protocol and technologies applied in this experiment enable TF QKD to achieve high secure key rate at long distribution distance, and hence practically useful for field implementation of intercity QKD.
\end{abstract}

\maketitle

{\it Introduction.---}
Channel loss seems to be the most severe limitation on the practical application of long distance quantum key distribution (QKD)~\cite{Gisin2002,Scarani2009,zhang2018large}, given that quantum signals cannot be amplified. Much efforts have been made towards the goal of a longer-distance for QKD ~\cite{liao2017satellite,MDI404km,boaron2018secure}. Theoretically, the decoy-state method~\cite{H03Quantum,wang05Beating,LMC05Decoy} can improve the key rate of coherent-state based QKD from scaling quadratically to a linear with the channel transmittance, as what behaves of a perfect single-photon source. This method can beat the photon-number-splitting attack to the imperfect single-photon source and the coherent state is used as if only those single-photon pulses were used for key distillation, and hence it can reach the key rate to a level comparable with that of a perfect single-photon source.

Remarkably, with the ideal of twin-field QKD(TF-QKD)~\cite{nature18Overcoming,wang2018sns,Ma2018PhaseMatching,tamaki2018information,Cui2019TwinField,Curty2018Simple,Yu2019Sending, Lin2018Simple}, the secure key rate can be further improved to the scale of the square root of the channel transmittance. This TF-QKD can also drastically improve secure distance of QKD. It shows that, the coherent-state source can actually be an advantage over the single-photon source because it can make use of the linear superposition of vacuum and one photon state of the twin field from Alice and Bob. This method has the potential to achieve a key rate that scales with the square root of channel transmittance, and can by far break the known distance records of existing protocols in practical QKD~\cite{MDI404km,boaron2018secure}. Although the theoretical secure key rate can be even higher, any key rate of a repeaterless QKD protocol is bounded by the repeaterless secret key capacities, known as the Takeoka-Guha-Wilde (TGW) bound~\cite{takeoka2014fundamental} and the Pirandola-Laurenza-Ottaviani-Bianchi (PLOB) bound~\cite{PLOB2017}.

So far, there are a number of experiments\cite{minder2019experimental,Liu2019Experimental,wang2019beating,zhong2019proof,fang2019surpassing} for TF-QKD, which broke the repeater-less QKD bound ~\cite{PLOB2017}, with~\cite{wang2019beating,Liu2019Experimental,fang2019surpassing} or without~\cite{minder2019experimental,zhong2019proof} real fiber, with~\cite{Liu2019Experimental,fang2019surpassing} or without~\cite{minder2019experimental,wang2019beating,zhong2019proof} considering the finite size effect.

However, we are still interested in the target of breaking the {\em absolute limit} of repeater-less QKD. The {\em absolute limit} here is the unconditional bound value for the repeater-less QKD given {\em whatever} devices including perfect detection devices. Breaking such a bound is meaningful because we do not have to request any device condition for the repeater-less QKD in comparison. This is different from breaking the relative bound which is calculated under the restriction of actual detection efficiency, a factor of value around 0.3 to the absolute limit. Meanwhile, besides beating the rate distance limit, chasing longer distance is always another major goal for QKD research. Here we experimentally demonstrate unconditionally breaking the limit of repeater-less QKD at various distances and extend the QKD secure distance to 509 km. Especially, the following properties of our experiment makes it merit the {\em unconditionally} result in the breakthrough: 1) Our setup uses two independent laser sources. There is no incidental light to Alice and Bob, and hence there is no need to monitor the incident light as the implementations which directly using seed light from Charlie. In this sense, our experiment keeps the property of measurement-device-independent for the security. 2)We have taken the finite key effect into consideration with a failure probability of $10^{-10}$. This makes the security based directly on the final key of the real experiment itself rather than a security on the imagined final key from infinite number of pulses. 3)Our results break the absolute key rate limit of repeater-less QKD and could be directly deployed in the field.

Since the absolute limit of the key rate is 3 to 4 times of the relative limit in the prior art experiment\cite{Liu2019Experimental}, we have to fully upgrade our whole system in both theory and experimental side.

First, we adopt the protocol of SNSTF QKD \cite{wang2018sns} with improved method of post data processing, in the name of AOPP protocol\cite{xu2019general} with finite size effects~\cite{jiang2019unconditional,Jiang2019Higher} being considered. Here, we implement a practical four-intensity method~\cite{Yu2019Sending} for decoy-state analysis, where each party exploits four different intensities, namely,  $0, \mu_1, \mu_2$ and $\mu_z$. To improve the key rate, we take  bit error rejection by active odd parity pairing(AOPP)\cite{xu2019general} in the post data processing stage. In this way, the sending probability in $Z$ basis can be far improved and hence largely improves the number of effective events. As a result, the final key rate is improved a lot especially in the case of small data size with finite key effects being considered. Detailed calculation of AOPP is presented in the supplement.

\begin{figure*}[tbh]
\centering
\resizebox{14cm}{!}{\includegraphics{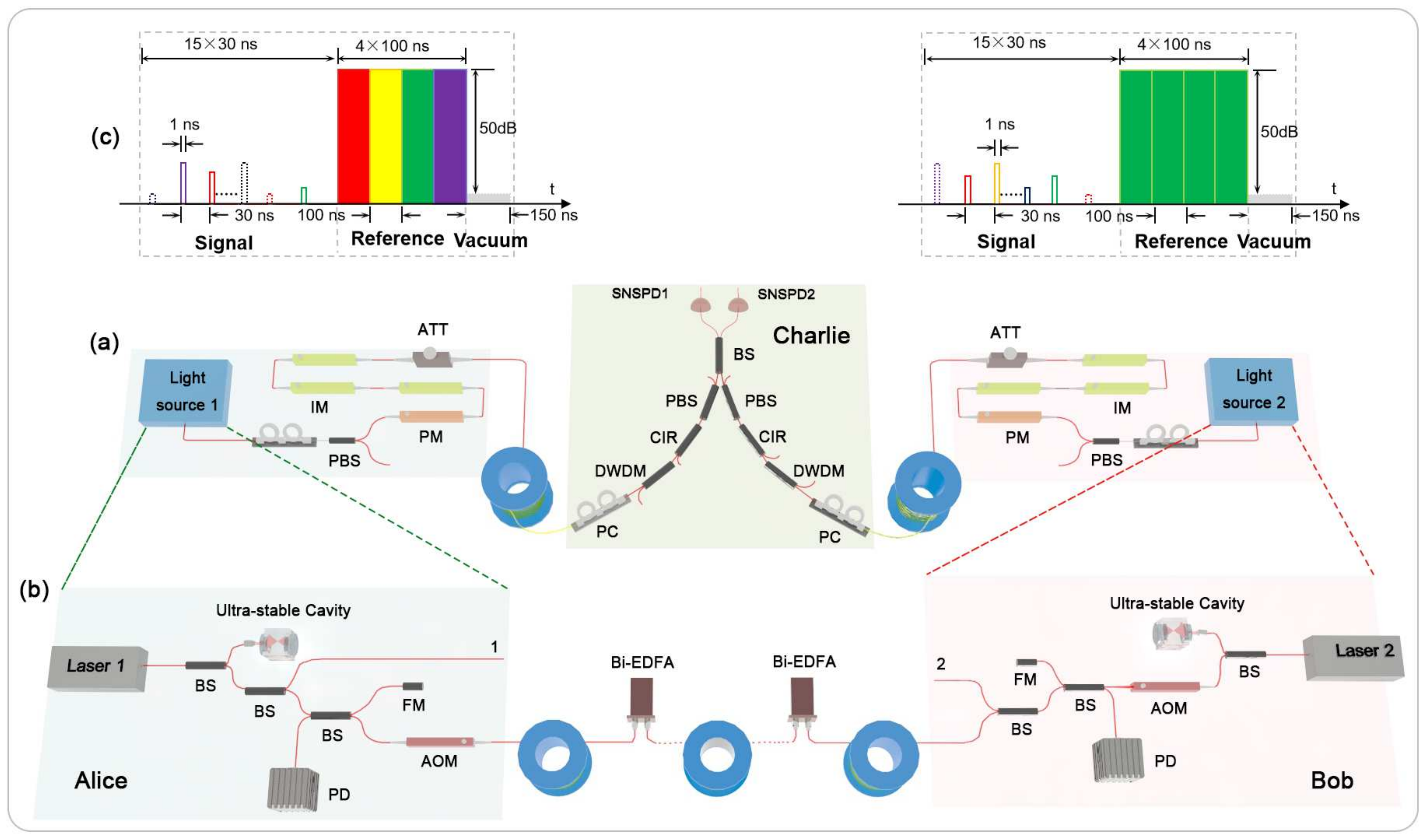}}
\caption{(a) Schematic of our experimental setup. Alice and Bob use remotely frequency-locked stable continuous wave (CW) lasers as sources. These light sources are then modulated by a phase modulator (PM) and three intensity modulators (IM1, IM2, IM3) for phase randomization, encoding, and decoy intensity modulation. The pulses are then attenuated by an attenuator (ATT) and sent out via fiber spools to Charlie. At Charlie's station, the encoded pulses from Alice and Bob are adjusted with polarization controllers (PCs) and polarized with polarization beam splitters (PBSs), then filtered with Dense Wavelength Division Multiplexings (DWDMs)  and circulators (CIRs), and finally interfered at a beam splitter (BS). The pulses are detected by superconducting nanowire single-photon detectors (SNSPDs). (b) Remote Frequency-locking system for Alice's and Bob's light sources. The fiber length fixed to 500 km for all experimental tests. 9 bi-directional erbium-doped fiber amplifiers (Bi-EDFAs) are deployed in the fiber path. BS: beam splitter, AOM: acousto-optic modulator, Bi-EDFA: bidirectional erbium-doped fiber amplifiers, FM: Faraday mirror, PD: photodiode. (c) The time sequence of the basic modulation period. During the basic modulation period of 1 $\mu$s, Alice (Bob) randomly modulate a intensity to $\mu_x$, ($\mu_x\in\{\mu_z, \mu_1, \mu_2, 0\}$) and phase to $\theta_A$ ($\theta_B$) ($\theta_A$, $\theta_B\in\{$0$, \pi/8, 2\pi/8 ... 15\pi/8\}$) on 15 signal pulses with a 1 ns duration and 29 ns interval in the first 450 ns, in the next 400 ns, Alice modulate the intensity to $\mu_{ref}$ and phase orderly to 0, $\pi/2$, $\pi$, $3\pi/2$, while Bob modulate the intensity to $\mu_{ref}$ and phase to $\pi$ on 4 reference pulses with a 100 ns duration, finally they modulate the intensity to $0$ and phase to $\pi$ on a vacuum state pulses in the last 150 ns as the recovery time for the SNSPDs.}
\label{Fig:setup}
\end{figure*}

Besides optimizing the protocol, we have also substantially improved the experimental setup which is shown in Fig.~\ref{Fig:setup}(a). The realization of TF QKD is challenging comparing to other QKD protocols, because the precisely control of the relative phase from independent lasers through long distance fiber links are required to ensure high quality interference in the measurement station. The phase difference, however, can be accumulated by any wavelength differences between the light sources, or by fast phase drift in the fiber link~\cite{nature18Overcoming}.

The wavelengths of two independent lasers are locked with the time-frequency dissemination technology~\cite{Liu2019Experimental}, as shown in Fig.~\ref{Fig:setup}(b). Alice uses a commercial continuous wave laser source that internally locked to her cavity, yielding a linewidth less than 1 Hz at 1550.0465 nm central wavelength. Bob locks a stable continuous wave laser source to his cavity using Pound-Drever-Hall (PDH) technique~\cite{drever1983laser,pound1946electronic}, yielding a linewidth of approximate 1 Hz at a central wavelength of 1550.0474 nm. The frequency difference between Alice's and Bob's ultra stable lasers is set to 112 Megahertz before further process for the QKD setup. The relative frequency drift is measured to be approximately 0.1 Hz$\cdot s^{-1}$.

Alice divides her light into two parts, one of which is used as her QKD laser source, and the other is sent to Bob as a wavelength reference. Bob, receives Alice's reference light and compensate the frequency difference using an acoustic-optic modulator (AOM). Then he splits his locked light into two parts, one is used as his QKD laser source, and the other is sent to Alice to compensate the phase noise in the fiber between them. This is done by Alice using an AOM at her output. The fiber distance between Alice and Bob for this frequency and phase locking is fixed to 500 km in all experimental tests, in order to match the longest distance QKD experiment. The total loss of the 500 km single mode fiber is measured as 98.47 dB. 9 bi-directional erbium-doped fiber amplifiers (Bi-EDFAs) are used in the path to amplify the signal for frequency and phase locking. We note that this setup is able to deploy in field experiment without major changes.

The phase drift in the QKD fiber links is compensated with strong phase reference pulses. In this method, Alice and Bob periodically send strong phase reference pulses to the measurement station, where the interference result is recorded and analyzed to calculate the relative phase between Alice's and Bob's fiber~\cite{Liu2019Experimental}. We note that to acquire the phase the signal pulses experience, there are a few requirements for the phase reference pulses: the wavelength must be set to the same as the signal; they must transmit exactly the same optical fiber; thus, the phase reference pulses are time-multiplexing with the signal pulse. The intensity of the phase reference pulses are set with enough photons received at the measurement station.
Thus the peak intensity in the long distance experiment is high and thus will inevitably induce additional noise to the measurement besides the detectors' dark counts. This will be discussed in detail later.

Then in order to avoid the PNS attack~\cite{H03Quantum}, USD attack~\cite{Dusek2000Unambiguous} and implement phase estimation, the two users individually encode the light to 16 different phase slices with a phase modulator (PM) and 5 different intensities with three intensity modulators (IMs). The largest intensity pulses are used as reference pulses for phase estimation, while the other 4 are used as the signal state pluses, strong decoy state pluses, weak decoy state pluses, and vacuum state pluses. In order to implement a more than 500 km TF QKD, the intensity ratio between the reference pulses and the vacuum decoy state signal should be higher than 50 dB in the pulse duration. This requires a huge dynamic range for the IM. Here, we achieved a stable intensity modulation by placing the IMs in a thick foam box to reduce the fluctuations of the ambient temperature and designing a reasonable modulation waveform pattern. We set the basic period to 1 $\mu$s with a time sequence as shown in Fig.~\ref{Fig:setup}(c), in which we send 15 signal pulses, each with a 1 ns pulse duration and 29 ns interval for the first 450 ns, 4 strong phase reference pulses, each with a 100 ns pulse duration and different phase in the next 400 ns, and the vacuum states as the recovery time for the superconducting nanowire single-photon detectors (SNSPDs) in the last 150 ns.

Then we attenuate signals from both sides into single photon level with passive attenuators. At the measurement station (Charlie), the two beams are interfered at a beam splitter (BS) and the interference results are detected by 2 SNSPDs and recorded by a high speed multi-channel time tagger. Both the signal and the reference pulses are detected by the SNSPDs, thus, low dark count rate, high detection efficiency and high count rate are required at the same time. We improve the SNSPD by integrating a filter on the end face of the coupling fiber to reduce the dark count and the insertion loss compared to that in~\cite{Liu2019Experimental}. A resistor is inserted in series to the SNSPD at low temperature to accelerate the recovery process and avoid the latching effect~\cite{Zhang2018FiberCoupled}. The dark count of the two SNSPDs are measured to be both less than 3.5 Hz, with the detection efficiencies of 56\% and 58\%. The maximum counting rate of the detectors are tested to be approximately 10 MHz with continuous light as input.

As mentioned previously, the reflection and scattering of the strong reference pulses will induce various of noise. The forward Rayleigh scattering might be the strongest effect in the long fiber. However, it would not influence the signal because it remains in the same time period as the reference pulses, which is at least 14.5 ns separation to the signal in time domain. It is the same case for Brillouin scattering noise with small frequency differency. Raman scattering noise has a ~THz frequency shift and the chromatic dispersion will move the noise into the signal time span. According to our calculation, the Raman scattering noise may induce about 100 cps at 250 km and will increase with larger intensity of the phase reference pulses. This confirms with our test result. We insert a 100G Dense Wavelength Division Multiplexing (DWDM) filter to eliminate this noise. When eliminating the above noises, there are still a few noise source beside the detector dark count: the reflection of SNSPD or the backward Rayleigh scattering will again introduce backward Rayleigh scattering noise that transmit in the same direction with the signal and distribute uniformly in time domain. The scattering noise from the reflection of SNSPD is about 100 cps and can be removed by inserting a circulator before the SNSPD. The Re-Rayleigh scattering of Rayleigh scattering (RRSORS) is, however inevitable in our experiment.

\begin{figure}[tbh]
\centering
\resizebox{9cm}{!}{\includegraphics{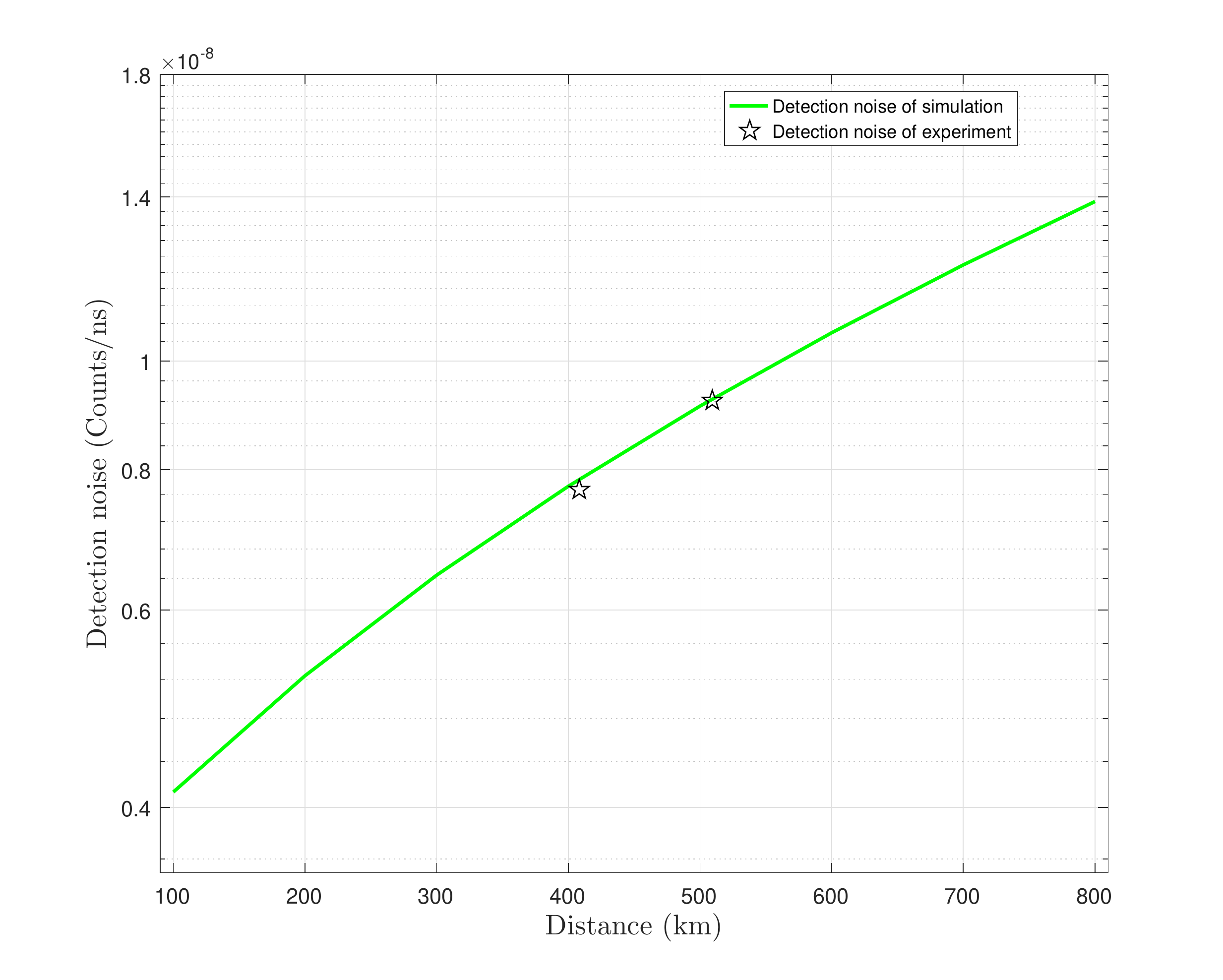}}
\caption{Theoretical and experimental noise rate with different fiber lengths. Alice and Bob are assumed to emit at the working intensity, with 2 MHz reference counts detected. The green curve is the theoretical simulation based on the Re-Rayleigh scattering of Rayleigh scattering (RRSORS) model and detector dark counts. The black stars are the experimental results.}
\label{Fig:noise}
\end{figure}

Here, we demonstrate a RRSORS model in fiber to estimate the detection noise by the following formula:
\begin{equation}
d=\frac{P_0S^2}{4{E_\nu}\alpha}e^{-{\alpha}l}[l+\frac{e^{-2{\alpha}l}}{2\alpha}-\frac{l}{2\alpha}]+D_c
\label{eq:detectionnoise}
\end{equation}

where $d$ is the detection noise count, $E_\nu$ is the photon energy, $l$ is the fiber length, $\alpha$ is the loss coefficient of fiber, $P_0$ is the intensity of light sent into the fiber, $D_c$ is the dark count of SNSPD. As shown in Fig.~\ref{Fig:noise}, we tested the detection noise in our experiment, which is almost agreement with the estimation results and increases with distance as the predominance of RRSORS noise over long distances. Fortunately, the RRSORS noise is at the same level as the SNSPD dark counts and acceptable at 500 km scale. Additionally, we tested the detection noise caused by RRSORS corresponding to different count rates with 250 km standard optical fiber, which is also almost agreement with the estimation results(see Supplemental Materials for details about the RRSORS model).

After all these upgrades, we performed SNS-TF-QKD with the 350 km standard optical fiber, 408 km and 509 km ultra-low-loss optical fiber between Alice and Bob. The corresponding detailed parameters which include the optical efficiencies of the fibers and optical devices at the measurement site, the proportions and intensities of each states for each fiber lengths are summarized in Supplemental Materials.

\begin{figure}[tbh]
\centering
\resizebox{9cm}{!}
{\includegraphics{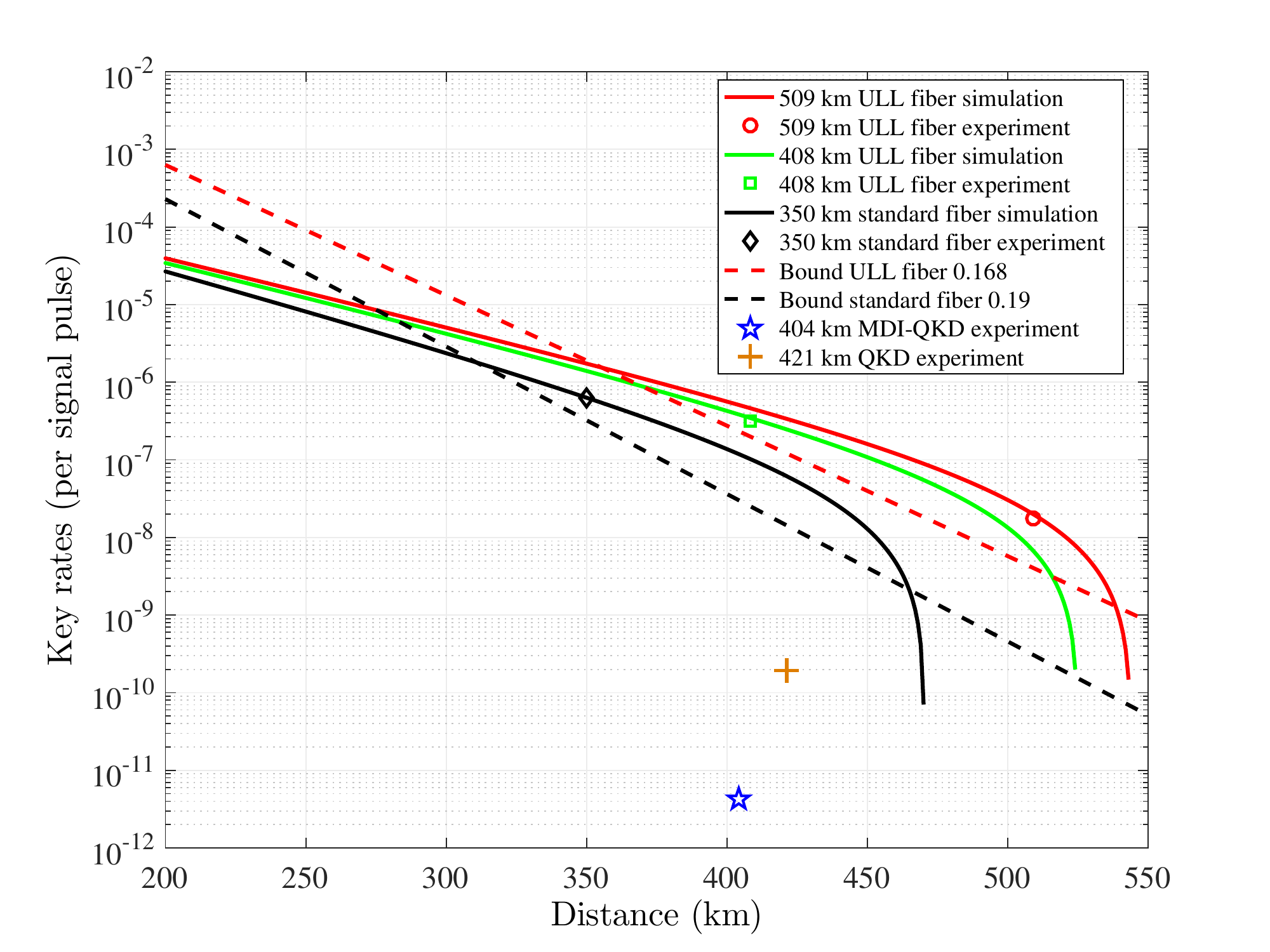}}
\caption{SNS-TF-QKD secure key rates. The black diamond point is the experimental secure key rate with 350 km standard single mode fiber; The green square point is the experimental secure key rate with 408 km ultra-low loss fiber; And the red circle point is the experimental secure key rate with 509 km ultra-low loss fiber. The orange cross point shows the experimental secure key rate of the three-state time-bin QKD~\cite{boaron2018secure}, and the blue star point shows the experimental secure key rate of 4-intensity decoy-state MDI-QKD~\cite{MDI404km}. The black, green and red curve respectively shows the simulation result for standard fiber length of 350 km, ultra-low loss fiber length of 408km and 509 km with noise probability of $1\times10^{-8}$ and X-basis baseline error of 0.04; Finally, the black and red dashed curve show the absolute key rate limit of repeaterless of standard fiber and ultra-low loss fiber.}
\label{Fig:bound}
\end{figure}

With the considering of the finite data size effect, we calculated the secure key rate~\cite{jiang2019unconditional}. The experimental results are presented in Fig.~\ref{Fig:bound}, the final key rates break the absolute linear bound at all distances. In the  350 km standard optical fiber experiment, the total pulses sent is $3\times 10^{11}$, the valid detections are $7.55\times10^6$, and the secure key rate is calculated to $R=6.34\times10^{-7}$. Similarly, with a total pulses $5.2\times 10^{11}$ sent in the 408 km experiment, the valid detections are $1.0\times10^7$, the secure key rate is $R=3.2\times10^{-7}$, which is more than 3 orders of magnitude higher than that reported in~\cite{boaron2018secure} and 4 orders of magnitude higher than that in~\cite{MDI404km}. Further, we perform an experiment with 509 km fiber distance. With $1.55\times10^{12}$ pulses sent, the valid detections are $2.76\times10^6$, and the secure key rate is $R=1.79\times10^{-8}$. This secure key rate is higher than the absolute PLOB bound in all these directions, this again verifies the high performance of the SNS-TF-QKD with AOPP.

In conclusion, we have developed remote optical frequency-locking technique, experimentally implemented the SNS-TF-QKD protocol with two independent lasers to break the absolute key rate limit of repeaterless over 350 km, 408 km and 509 km, and demonstrated a RRSORS model to verify the predominance of the detection noise over long distances. Moreover, the 500 km long accompanying fiber link for frequency lock makes our system naturally fit in a field test, which will be the next step for this research.

Another interesting question for the future research is how to approach the limit of TF QKD. Currently, the dominated noise is RRSORS within fiber, which is inevitable with our design. One possible solution is to exploit optical frequency comb instead of strong reference pulses. The optical frequency comb can simultaneously emit multiple different wavelength continuous wave lasers while maintain stable phase difference between each other~\cite{Udem2002optical,Holzwarth2000Optical}. This characteristic of the optical frequency comb is expected to achieve a different wavelength modulation of the reference pulses and the signal pulses in SNS-TF-QKD to avoid the RRSORS.

\emph{Acknowledgements.---}

This work was supported by the National Key R\&D Program of China (2017YFA0303901, 2017YFA0304000), the National Natural Science Foundation of China, the Chinese Academy of Science, the Anhui Initiative in Quantum Information Technologies, the Shanghai Sailing Program.

\bibliography{BibSNSTFQKD-1003}

\newpage

\section{Theory of improved SNS-TF-QKD protocol}
In our experiment, the 4 intensity SNS protocol with AOPP proposed in ~\cite{xu2019general,Jiang2019Higher} is used. In the protocol, Alice and Bob will repeat the following process for $N$ times to obtain a series of data:
\noindent In each time window, Alice (Bob) randomly decides whether it is a decoy window with probability $1-p_z$, or a signal window with probability $p_z$. If it is a signal window, with probability $\varepsilon$, Alice (Bob) randomly prepares a phase-randomized weak coherent state (WCS) pulse with intensity $\mu_z$, and denote it as bit $1$ ($0$); with probability $1-\varepsilon$, Alice (Bob) prepares a vacuum pulse, that is, doing nothing and denote it as bit $0$ ($1$). If it is a decoy window, Alice (Bob) randomly prepares a pulse of coherent state $\ket{0}$, $\ket{e^{i\theta_A}\sqrt{\mu_{1}}}$ or $\ket{e^{i\theta_A^\prime}\sqrt{\mu_{2}}}$ (state $\ket{0}$, $\ket{e^{i\theta_B}\sqrt{\mu_{1}}}$ or $\ket{e^{i\theta_B^\prime}\sqrt{\mu_{2}}}$) with probabilities $p_{0}$, $p_{1}$ and $1-p_{0}-p_{1}$, respectively, where $\theta_A,\theta_A^\prime,\theta_B$ and $\theta_B^\prime$ are different in different windows, and are random in $[0,2\pi)$. Then Alice and Bob send their prepared pulses to Charlie. Charlie is assumed to perform interferometric measurements on the received pulses and announces the measurement results to Alice and Bob. If one and only one detector clicks in the measurement process, Charlie also tells Alice and Bob which detector clicks, and Alice and Bob take it as an one-detector heralded event.

Then Alice and Bob announce the basis they used in each time window in the public channel. For a time window where both of Alice and Bob used the signal window, it is a $Z$ window. And the one-detector heralded events in $Z$ windows are called effective events. Alice and Bob get two $n_t$-bit strings $Z_A$ and $Z_B$ formed by the corresponding bits of effective events of $Z$ windows. Strings $Z_A$ and $Z_B$ will be used to extract the secure final keys. The intensity of pulses, that is the decision of sending or not sending in $Z$ windows are kept private, but the intensities of other pulses would be publicly announced after Alice and Bob finished basis calibration. For a time window where both of Alice and Bob used the decoy window and the intensities of WCS pulses they sent out are the same, Alice and Bob also announces the phase information $\theta_A,\theta_A^\prime,\theta_B$ and $\theta_B^\prime$ in the public channel. And if the phases of WCS pulses satisfy
\begin{equation}
1-\vert \cos(\theta_A-\theta_B-\psi_{AB})\vert\le \lambda,
\end{equation}
or
\begin{equation}
1-\vert \cos(\theta_A^\prime-\theta_B^\prime-\psi_{AB})\vert\le \lambda,
\end{equation}
it is an $X$ window. Here $\psi_{AB}$ can take an arbitrary value which can be different from time to time as Alice and Bob like, so as to obtain a satisfactory key rate for the protocol~\cite{Liu2019Experimental}. $\lambda$ is a positive value close to $0$, and would be optimized to obtain the highest key rate. And the one-detector heralded events in $X$ windows are called effective events.

We have the following definitions to clearly show our calculation process. We denote $\rho_0=\oprod{0}{0}, \rho_1=\sum_{k=0}\frac{\mu_1^ke^{-\mu_1}}{k!}\oprod{k}{k}, \rho_2=\sum_{k=0}\frac{\mu_2^ke^{-\mu_2}}{k!}\oprod{k}{k}$ and $\rho_z=\sum_{k=0}\frac{\mu_z^ke^{-\mu_z}}{k!}\oprod{k}{k}$, where $\rho_1$ and $\rho_2$ are the density operators of the phase-randomized WCS used in decoy windows and $\rho_z$ is the density operator of the phase-randomized WCS used in the signal window. And this also applies to Bob's quantum state. In the whole process, Alice and Bob obtain $N_{\alpha\beta}(\alpha\beta=\{00,01,02,10,20\})$ instances when Alice sends state $\rho_\alpha$ and Bob sends state $\rho_\beta$. And after Charlie's measurement, Alice and Bob obtain $n_{\alpha\beta}$ one-detector heralded events. We denote the counting rate of source $\alpha\beta$ as $S_{\alpha\beta}=n_{\alpha\beta}/N_{\alpha\beta}$, and its corresponding expected value as $\mean{S_{\alpha\beta}}$.

For the original SNS protocol, the lower bounds of the counting rate of untagged bits, $\underline{\mean{s_{10}}}$ and $\underline{\mean{s_{01}}}$, and the upper bound of its phase-flip error rate $\overline{\mean{e_{1}^{ph}}}$ are needed to calculate the secure final key rate. Here the untagged bits are caused by the effective events of $Z$ windows where only one party of Alice and Bob chooses to send out a phase-randomized WCS pulse and she or he actually sends out a single photon. All the one-detector heralded events except the effective events of the $Z$ windows are used to calculate $\underline{\mean{s_{10}}}$ and $\underline{\mean{s_{01}}}$, which are~\cite{Jiang2019Higher}
\begin{align}
\label{s01mean}\underline{\mean{s_{01}}}&= \frac{\mu_{2}^{2}e^{\mu_{1}}\underline{\mean{S_{01}}}-\mu_{1}^{2}e^{\mu_{2}}\overline{\mean{S_{02}}}-(\mu_{2}^{2}-\mu_{1}^{2})\overline{\mean{S_{00}}}}{\mu_{2}\mu_{1}(\mu_{2}-\mu_{1})},\\
\underline{\mean{s_{10}}}&= \frac{\mu_{2}^2e^{\mu_{1}}\underline{\mean{S_{10}}}-\mu_{1}^2e^{\mu_{2}}\overline{\mean{S_{20}}}-(\mu_{2}^2-\mu_{1}^2)\overline{\mean{S_{00}}}}{\mu_{2}\mu_{1}(\mu_{2}-\mu_{1})},
\end{align}
where $\underline{\mean{S_{jk}}}$ and $\overline{\mean{S_{jk}}}$ are the lower and upper bounds estimated by its corresponding observed values with the help of Chernoff bound.

If both of Alice and Bob choose to send the WCS pulses with intensity $\mu_1$ in an $X$ window, we call it $X_1$ window. We denote the number of instances of $X_1$ windows as $N_{X_1}$. And for the effective event of the $X_1$ window that the phases of the pulse pair satisfy $\cos(\theta_A-\theta_B)> 0$ and Charlie announces that this pulse pair causes the right detector clicking, or the phases of the pulse pair satisfy $\cos(\theta_A-\theta_B)< 0$ and Charlie announces that this pulse pair causes the left detector clicking, it is a wrong effective event. We denote the number of wrong effective event of $X_1$ windows as $m_{X_1}$, and we have $T_{X_1}=m_{X_1}/N_{X_1}$, whose corresponding expected value is $\mean{T_{X_1}}$. A phase-flip error occurs when the one-detector heralded event in $X_1$ windows is a wrong effective event, thus we have~\cite{Jiang2019Higher}
\begin{equation}\label{e1}
\overline{\mean{e_1^{ph}}}=\frac{\overline{\mean{T_{X_1}}}-e^{-2\mu_{1}}\underline{\mean{S_{00}}}/2}{2\mu_1e^{-2\mu_1}\underline{\mean{s_1}}},
\end{equation}
where $\underline{\mean{s_1}}=(\underline{\mean{s_{01}}}+\underline{\mean{s_{10}}})/2$.

Before we use $\underline{\mean{s_{10}}},\underline{\mean{s_{01}}}$, and $\overline{\mean{e_{1}^{ph}}}$ to calculate the final key rate and extract the secure final keys, we can perform the actively odd-parity pairing (AOPP) on Alice's and Bob's strings to improve the final key rate.

\begin{figure}[tbh]
\centering
\resizebox{9cm}{!}
{\includegraphics{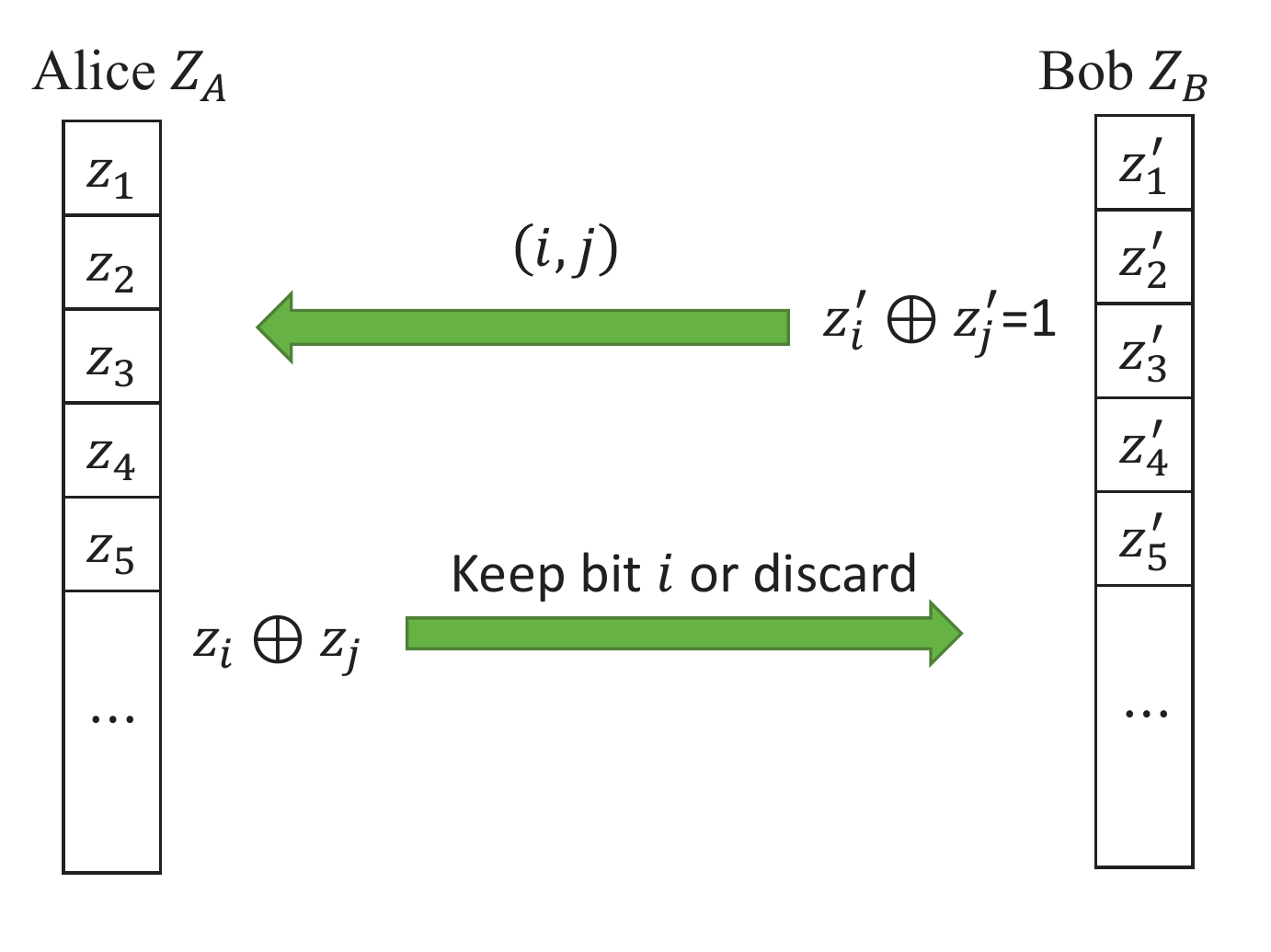}}
\caption{Flow chart of AOPP execution.}
\label{AOPP}
\end{figure}

AOPP is a pre-error correction process with two way communication. While performing AOPP, as shown in the schematic Fig.~\ref{AOPP}, Bob randomly pairs the bits in $Z_B$, where one bits are randomly chosen from all the left bits $0$, the other bit is randomly chosen from all the left bits $1$, and obtains $n_p=\min(n_{t0},n_{t1})$ pairs, where $n_{t0}$ is the number of bits $0$ and $n_{t1}$ is the number of bits $1$ in $Z_B$. The unpaired bits would be directly discarded. We denote the pair that contains $i$-th and $j$-th bit of $Z_B$ as $(z_i^\prime,z_j^\prime)$, where $z_i^\prime$ and $z_j^\prime$ are the corresponding bit values. Then Bob broadcasts all the location of bits in each pair $(z_i^\prime,z_j^\prime)$ to Alice. Alice calculates the value of $z_i\oplus z_j$ and broadcast it to Bob, where $z_i$ is the value of $i$-th bit and $z_j$ is the value of $j$-th bit of $Z_A$. If $z_i\oplus z_j\neq 1$, Alice and Bob discard this bit pair; If $z_i\oplus z_j= z_i^\prime\oplus z_j^\prime$, Alice and Bob keep the second bit of the bit pair. The remained bits form two new $n_t^\prime$-bit strings $Z_A^\prime$ and $Z_B^\prime$, which would be used to extract the secure final keys.
Although the bit flip error rate of strings $Z_A^\prime$ and $Z_B^\prime$, $E_Z^\prime$, is an observed value in experiment, we can estimate this value in theory to show the power of AOPP intuitively. Only the following two types of pair can be survived after AOPP:

\noindent \emph{1. The two bits in the pair are both caused by the events that only one of Alice and Bob decides sending a phase-randomized WCS pulse. And we denote the number of those pairs as $n_{cc}$.}

\noindent 2. \emph{One of the bit in the pair is caused by the event that both of Alice and Bob decide not sending, and the other bit in the pair is caused by the event that both of Alice and Bob decide sending. We denote the number of those pairs as $n_{vd}$.}

\noindent The bit flip error rate $E_Z^{\prime}$ is expected to be
\begin{equation}
E_Z^{\prime}=\frac{n_{vd}}{n_{cc}+n_{vd}}.
\end{equation}
As the counting rate of the events that both Alice and Bob decide not sending is pretty small, this process can reduce the bit flip error rate significantly and hence a large $\varepsilon$ can be used to improve the number of effective events, $n_t$, in the $Z$ windows. With the improvement of $n_t$, the final key rate can be improved by one or two times under the situation that the AOPP process drops at least half of the bits in $Z_A$ and $Z_B$. Note that the traditional error correction with two way communication can not improve the key rate under practical sources and channels~\cite{xu2019general}.

At same time as the parity of the pairs is always odd, one extra bit information of each pair has been leaked, thus the untagged bits now are the bits that survived from those bit pairs formed by two untagged bits. And we can estimate the lower bound of the number of those new untagged bits, $\underline{\mean{n_1^\prime}}$, according to $\underline{\mean{s_{01}}}$ and $\underline{\mean{s_{10}}}$, which is
\begin{equation}
\underline{\mean{n_1^\prime}}=\frac{\underline{\mean{n_{01}}}}{n_{t0}}\frac{\underline{\mean{n_{10}}}}{n_{t1}}n_p,
\end{equation}
where
\begin{align}
\underline{\mean{n_{10}}}=Np_{z}^2\epsilon(1-\epsilon)\mu_{z}e^{-\mu_{z}}\underline{\mean{s_{10}}},\\
\underline{\mean{n_{01}}}=Np_{z}^2\epsilon(1-\epsilon)\mu_{z} e^{-\mu_{z}}\underline{\mean{s_{01}}}.
\end{align}
The phase-flip error rate of those new untagged bits is $\mean{\overline{e_1^{ph\prime}}}=2\mean{\overline{e_1^{ph}}}(1-\mean{\overline{e_1^{ph}}})$~\cite{xu2019general}.

We can use the Chernoff bound to help us estimate the real value of the number of new untagged bits and the corresponding phase flip error rate, which are
\begin{equation}\label{n11e11}
n_1^\prime=\varphi^L(\mean{\underline{n_1^\prime}}),\quad e_1^{ph\prime}=\frac{\varphi^U(\mean{\underline{n_1^\prime}}\mean{\overline{e_1^{ph\prime}}})}{\mean{\underline{n_1^\prime}}},
\end{equation}
where $\varphi^L(x)$ and $\varphi^U(x)$ are defined in Eqs.~\eqref{observ}-\eqref{endd}.

Finally we can calculate the length of secure final keys by the following formula
\begin{equation}\label{r2}
\begin{split}
l_A=&n_1^\prime[1-h(e_1^{ph\prime})]-fn_t^\prime h(E_Z^\prime)-\log_2{\frac{2}{\varepsilon_{cor}}}\\
&-2\log_2{\frac{1}{\sqrt{2}\varepsilon_{PA}\hat{\varepsilon}}},
\end{split}
\end{equation}
where $h(x)=-x\log_2(x)-(1-x)\log_2(1-x)$ is the Shannon entropy.

With the formula of Eq.~\eqref{r2}, the protocol is $\varepsilon_{tol}$-secure, and $\varepsilon_{tol}=\varepsilon_{cor}+\varepsilon_{sec}$, where $\varepsilon_{sec}=2\hat{\varepsilon}+4\bar{\varepsilon}+\varepsilon_{PA}+\varepsilon_{n_1^\prime}$. Here, $\varepsilon_{cor}$ is the upper bound of probability that the strings of Alice and Bob are not same after error correction; $\bar{\varepsilon}$ is the upper bound of probability that the real value of phase-flip error rate of untagged bits is larger than $e_1^{ph\prime}$; $\varepsilon_{n_1^\prime}$ is the upper bound of probability that the real value of the number of new untagged bits is smaller than $n_1^\prime$, and $\varepsilon_{PA}$ is the failure probability of privacy amplification.  And we set $\varepsilon_{cor}=\hat{\varepsilon}=\varepsilon_{PA}=\xi$, besides, we have $\bar{\varepsilon}=3\xi$ and $\varepsilon_{n_1^\prime}=6\xi$ if we set the failure probability of Chernoff bound as $\xi$, thus the security coefficients, $\varepsilon_{tol}=22\xi=2.2\times10^{-9}$.
\section{The Chernoff bound}
Let $X_1,X_2,\dots,X_n$ be $n$ random samples, detected with the value 1 or 0, and let $X$ denote their sum satisfying $X=\sum_{i=1}^nX_i$. $\phi$ is the expected value of $X$. We have
\begin{align}
\label{mul}\phi^L(X)=&\frac{X}{1+\delta_1(X)},\\
\label{muu}\phi^U(X)=&\frac{X}{1-\delta_2(X)},
\end{align}
where we can obtain the values of $\delta_1(X)$ and $\delta_2(X)$ by solving the following equations
\begin{align}
\label{delta1}\left(\frac{e^{\delta_1}}{(1+\delta_1)^{1+\delta_1}}\right)^{\frac{X}{1+\delta_1}}&=\frac{\xi}{2},\\
\label{delta2}\left(\frac{e^{-\delta_2}}{(1-\delta_2)^{1-\delta_2}}\right)^{\frac{X}{1-\delta_2}}&=\frac{\xi}{2},
\end{align}
where $\xi$ is the failure probability. Thus we have
\begin{equation}\label{sjklower}
\phi^L({N_{\alpha\beta}S_{\alpha\beta}})=N_{\alpha\beta}\underline{\mean{S_{\alpha\beta}}},\phi^U({N_{\alpha\beta}S_{\alpha\beta}})=N_{\alpha\beta}\overline{\mean{S_{\alpha\beta}}}.
\end{equation}

Besides, we can use the Chernoff bound to help us estimate their real values from their expected values. Similar to Eqs.~\eqref{mul}- \eqref{delta2}, the observed value, $\varphi$, and its expected value, $Y$, satisfy
\begin{align}
\label{observ}&\varphi^U(Y)=[1+\delta_1^\prime(Y)]Y,\\
&\varphi^L(Y)=[1-\delta_2^\prime(Y)]Y,
\end{align}
where we can obtain the values of $\delta_1^\prime(Y)$ and $\delta_2^\prime(Y)$ by solving the following equations
\begin{align}
\left(\frac{e^{\delta_1^\prime}}{(1+\delta_1^\prime)^{1+\delta_1^\prime}}\right)^{Y}&=\frac{\xi}{2},\\
\label{endd}\left(\frac{e^{-\delta_2^\prime}}{(1-\delta_2^\prime)^{1-\delta_2^\prime}}\right)^{Y}&=\frac{\xi}{2}.
\end{align}

\section{Controlling the Relative Phase between Alice and Bob}
\subsection{Relative Phase Drift between Alice and Bob with independent sources}
In the experiment of SNS-TF-QKD, it's particularly significant to precisely control the relative phase of the twin fields which interfere on Charlie's beam splitter~\cite{Liu2019Experimental}. The differential phase fluctuations between Alice and Bob are subject to the wavelength difference of the two light sources and the relative phase drift of the different signal fiber links~\cite{nature18Overcoming}. Here, to compensate the wavelength difference of the two light sources, we employed the time-frequency dissemination technology to remotely lock the two independent lasers to be the same frequency. Then, in order to accurately compensate for the relative phase drift of different signal fiber links, periodically, the two users transmitted phase reference pulses to Charlie, while Charlie recorded the detections of the interference in a limited cumulative time with the finite intensity of reference pulses to estimate the most possible relative phase for a post processing~\cite{Liu2019Experimental}. The cumulative time of reference detections needs to be set to get enough reference counts for an accurate estimation of the relative phase, is mainly subject to the relative phase drift rate of the twin-field over long distances. Therefore, with the remotely locked light sources, we tested the relative phase drift of all the different distances with two same length signal fiber links.

\begin{figure*}[tbh]
\centering
\resizebox{14cm}{!}
{\includegraphics{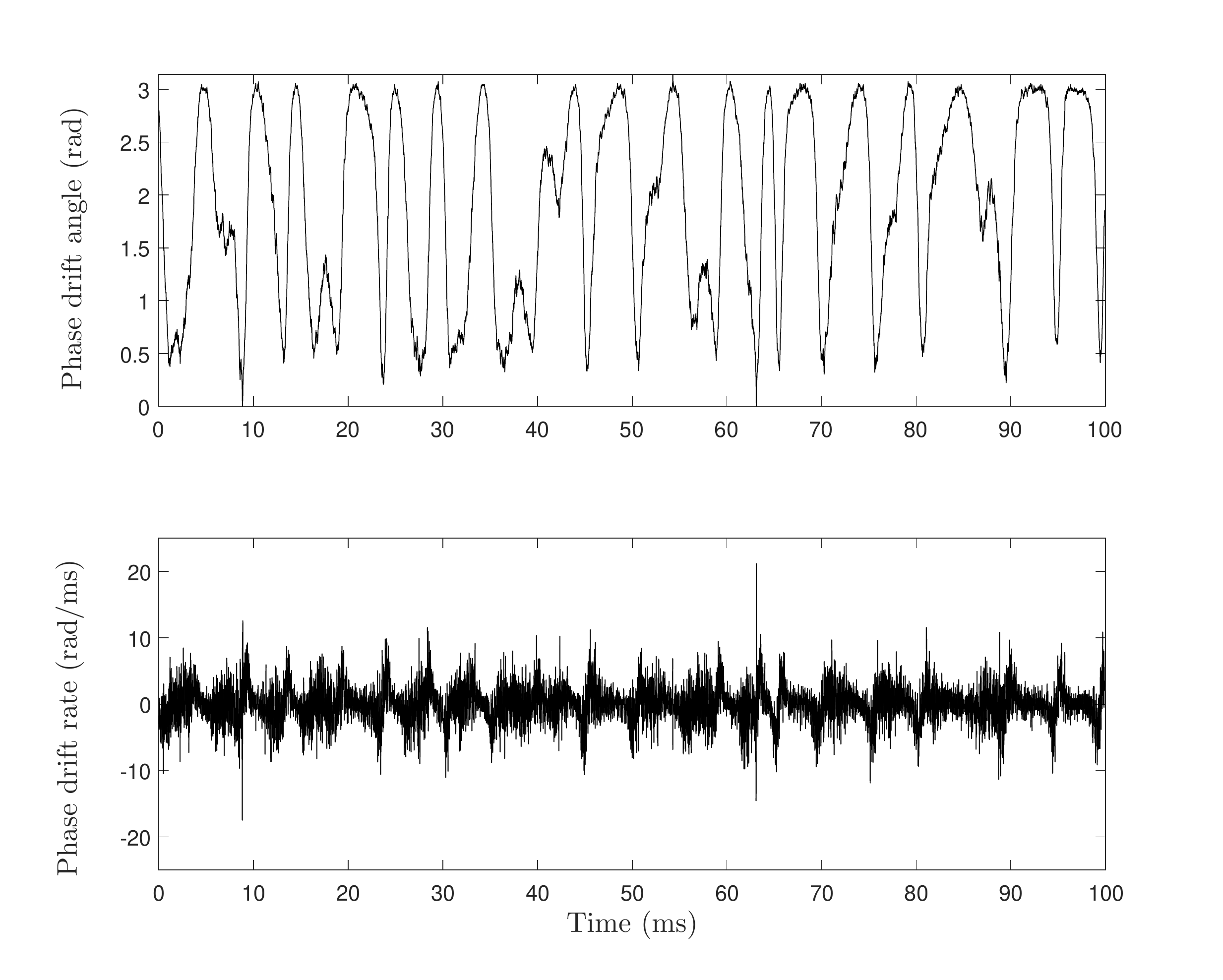}}
\caption{The relative phase drift of 0 km total signal fiber links.}
\label{Fig:0km}
\end{figure*}

The result of the relative phase drift (phase drift angle and the phase drift rate) with 0 km total signal fiber links is shown in Fig.~\ref{Fig:0km}, the corresponding phase drift rate follows a Gaussian distribution with a standard deviation of $2.72 rad\cdot ms^{-1}$.

\begin{figure*}[tbh]
\centering
\resizebox{14cm}{!}
{\includegraphics{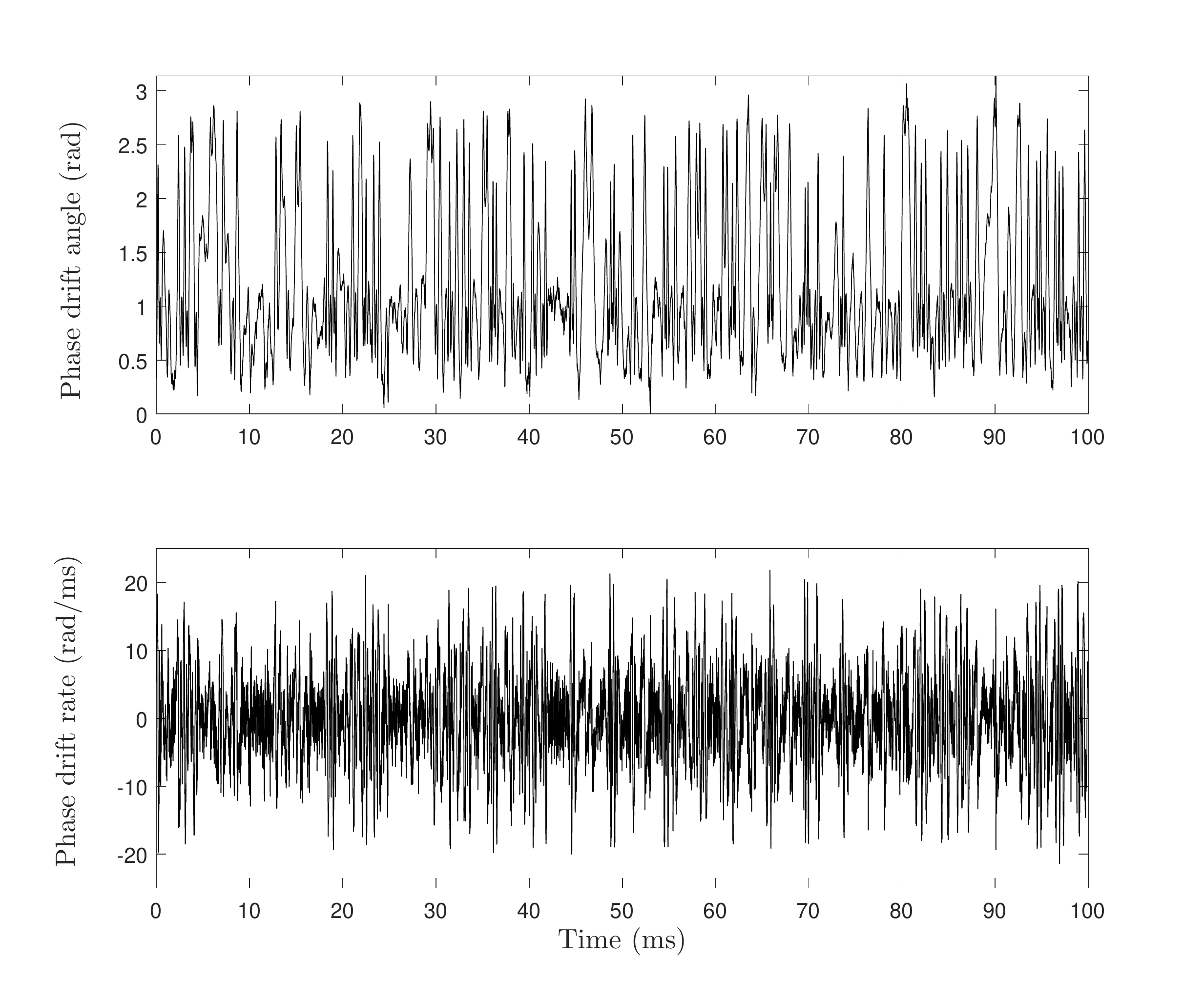}}
\caption{The relative phase drift of 350 km total signal fiber links.}
\label{Fig:350km}
\end{figure*}

Following the 0km, we tested the total signal fiber links of 350 km, the phase drift result is shown in Fig.~\ref{Fig:350km}, and the phase drift rate follows a Gaussian distribution with a standard deviation of $6.89 rad\cdot ms^{-1}$.

\begin{figure*}[tbh]
\centering
\resizebox{14cm}{!}
{\includegraphics{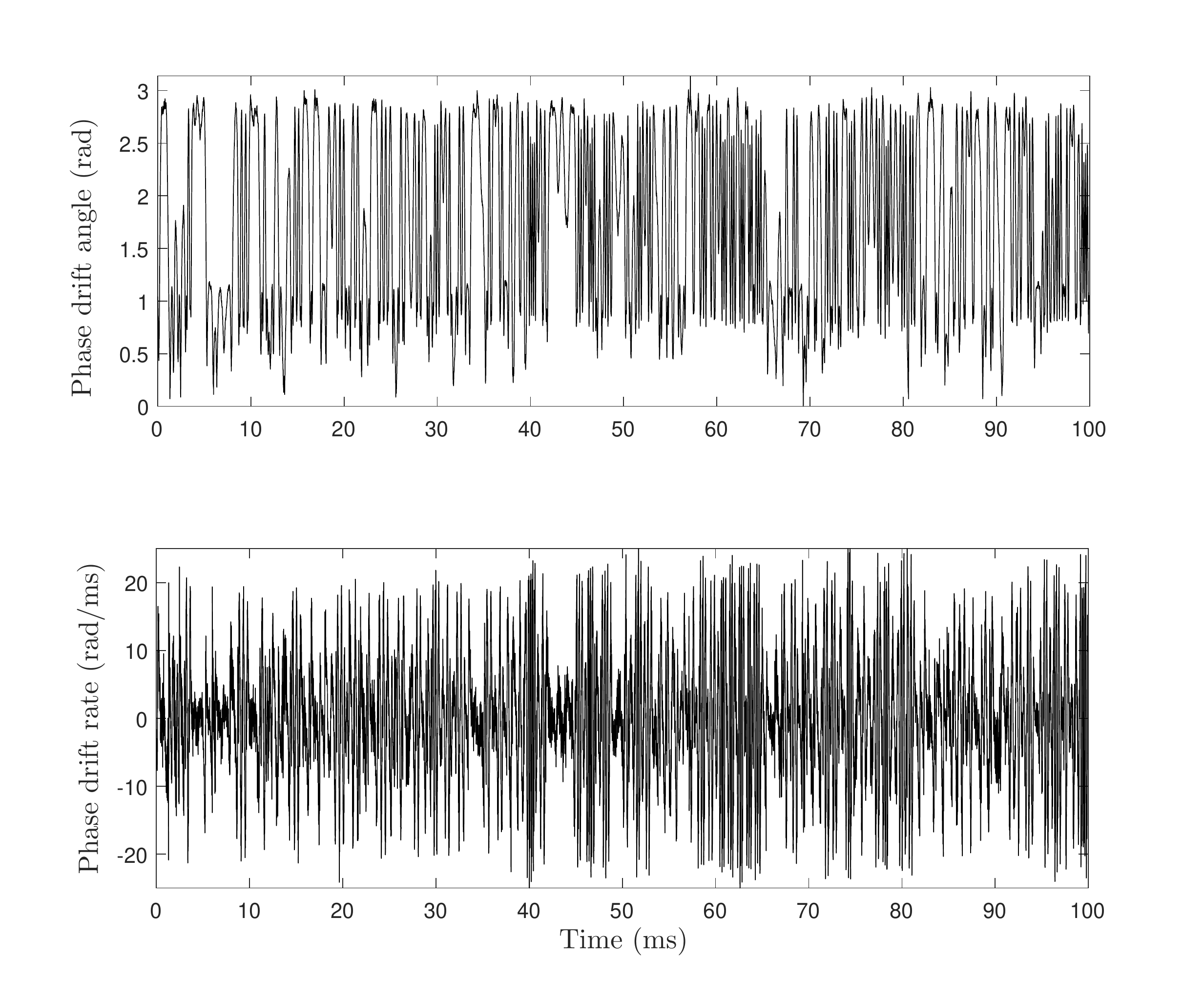}}
\caption{The relative phase drift of 408 km total signal fiber links.}
\label{Fig:408km}
\end{figure*}

\begin{figure*}[tbh]
\centering
\resizebox{14cm}{!}
{\includegraphics{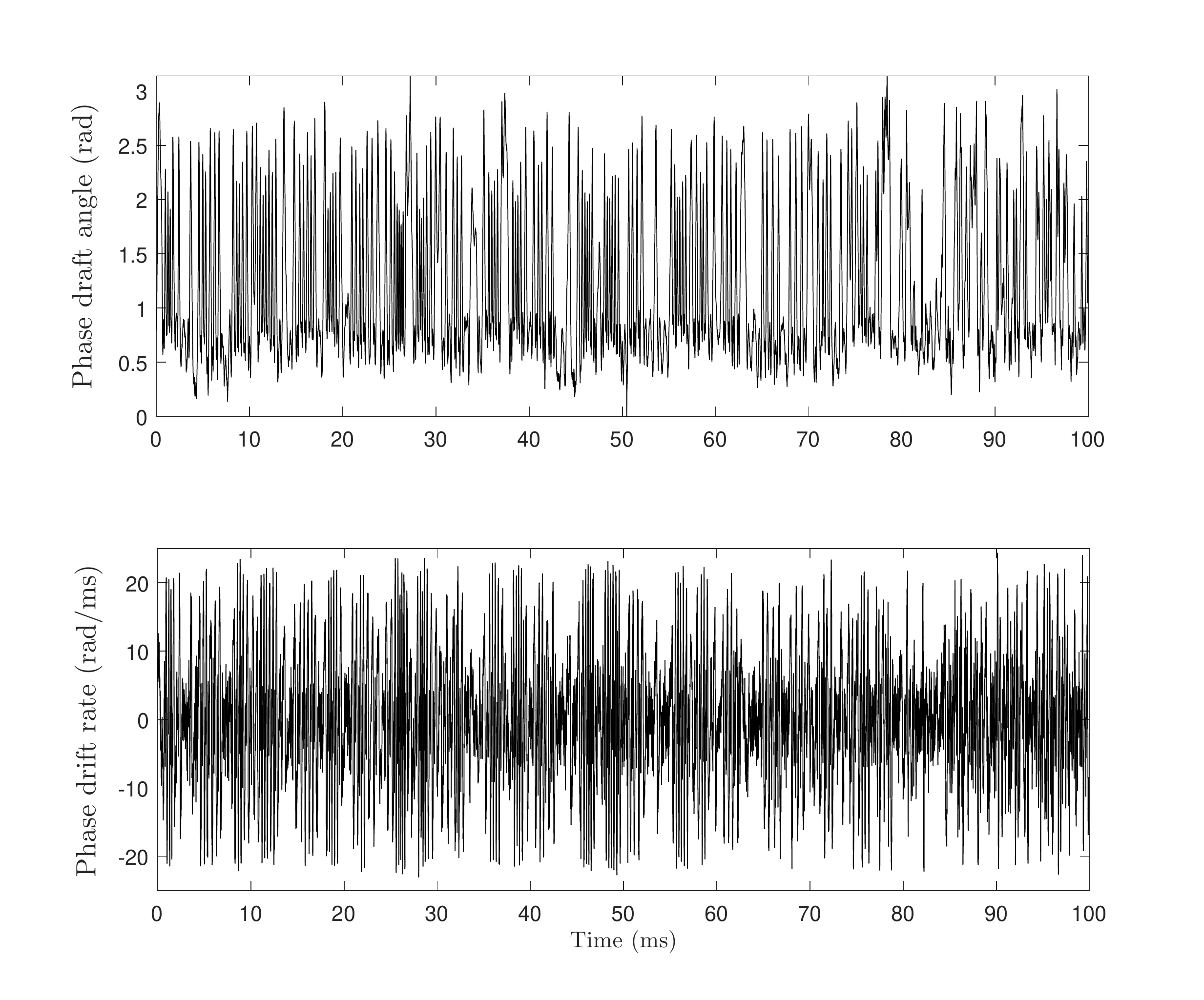}}
\caption{The relative phase drift of 509 km total signal fiber links.}
\label{Fig:509km}
\end{figure*}

Furthermore, we extended the total signal fiber links to 408 km and 509 km, the corresponding relative phase drift results are shown in Fig.~\ref{Fig:408km} and Fig.~\ref{Fig:509km}, the respective phase drift rates follows a Gaussian distribution with the standard deviation of $9.52 rad\cdot ms^{-1}$ and $9.58 rad\cdot ms^{-1}$.

In our experiment, the standard deviation of the relative phase drift rate is less than 9.6 $rad\cdot ms^{-1}$ for all scenarios, even over the 509 km signal fiber link. According to the maximum phase drift rate of 509 km ($25 rad\cdot ms^{-1}$), we applied accumulating of 12 basic periods of the reference detections for the relative phase estimation, the corresponding relative phase drift is around 0.3 rad (or 17 degree) which contributes to an error is less than $4\%$ when Alice and Bob sent the same phase.

\subsection{Estimating the Relative Phase Drifts of the twin-field}
In the experiment, instead of compensating for the relative phase drift in real-time with active feedback, the two users periodically transmitted phase reference pulses to Charlie, and Charlie recorded the interference results for estimating the most possible relative phase of the twin-field~\cite{Liu2019Experimental}.

As mentioned in the subsection of encoding details, Alice and Bob send 4 reference pulses to Charlie with 400 ns duration in one period. One after another, Alice modulates the phase to 0, $\pi/2$, $\pi$, $3\pi/2$, while Bob modulates the phase to $\pi$ on 4 reference pulses with a each 100 ns pulse duration, and Charlie successively records the interference detections of the 4 pairs of pulses. After several periods of accumulation, Charlie counts the total detections of the two detectors of each phase difference as $N_{i}$, where $i=1...4$ indicates the different phase difference between Alice and Bob are
\{$\pi$, $3\pi/2$, $0$, $\pi/2$\}.
Then he calculate the normalized counts probabilities as:
\begin{equation}
p_{i}=2 N_{i}/\Sigma N_{i}
\label{Eq:Prop}
\end{equation}
Similarly, these theoretical probabilities are calculated as:
\begin{equation}
p_{Ti}(\Delta\varphi_T)=\cos^2(\frac{\Delta\theta_i+\Delta\varphi_T}{2})
\end{equation}
where $p_{Ti}(\Delta\varphi_T)$ is the theoretical detection probability when the relative phase of the twin-field is $\Delta\varphi_T$, and the phase differences are $\Delta\theta_i=$\{$\pi$, $3\pi/2$, $0$, $\pi/2$\}, for $i=1...4$.
Finally, he minimize the error model $Err(\Delta\varphi_T)$ between counts and theoretical probabilities to estimate $\Delta\varphi_T$~\cite{Liu2019Experimental}.
\begin{equation}
Err(\Delta\varphi_T) = \sum_i{[p_i-p_{Ti}(\Delta\varphi_T)]^2}
\label{Eq:Err}
\end{equation}

It's a challenge to accurately estimate the relative phase that the finite intensity of the reference pulses and the limited cumulative time of reference detections which is subject to the relative phase drift rate of the twin-field over long distances. In our experiment, we optimized the numbers of the period to 12 to accumulate the counts of the two detectors for the relative phase estimation, which means that the statistics counted up to 12 us, the corresponding total counts of the 4 reference pulses are about 40.  And we also tested the total counts about 25 is almost the minimum counts for an accurate phase estimation.

\subsection{Compensating for the relative phase via post-processing}
In the theoretical analysis, after successful phase estimation, Charlie needs to stabilize the relative phase to 0 by a phase compensation. Here, instead of actively compensating, we apply a reasonable post selection of $X$-basis events to estimate the error rate in $X$-basis by the phase slice criterion~\cite{Liu2019Experimental}.
\begin{equation}
1-|\cos(\theta_A-\theta_B+\Delta\varphi_T)|<\Lambda
\end{equation}
with the estimated values of $\Delta\varphi_T$, the corresponding value of the $\Lambda$ is set to 0.015 in the experiment.

\section{ Encoding details of the Experiment}
In the experiment, with pre-generated quantum numbers, Alice and Bob implement the modulation by controlling an arbitrary-wave generator of which the sampling rate is 2 GHz with 14-bit depth and the maximum amplitude of output signal is about 500 mV. Then the two users amplify the modulation signal by more than 25 dB to drive the modulators to encode. With a basic encoding period of 1 $\mu$s, Alice (Bob) encodes the light to 16 different phase slices with a phase modulator (PM) and 5 different intensities with three intensity modulators (IMs). For the phase encoding, during the basic period of 1 $\mu$s, Alice (Bob) randomly modulates the phase to $\theta_A$ ($\theta_B$) on 15 signal pulses with a 30 ns duration in the first 450 ns, $\theta_A$, $\theta_B\in\{0, \pi/8, 2\pi/8 ... 15\pi/8\}$, in the next 400 ns, Alice orderly modulates the phase to 0, $\pi/2$, $\pi$, $3\pi/2$, while Bob modulates the phase to $\pi$ on 4 reference pulses with a each 100 ns duration, and then they both modulate the phase to $\pi$ on vacuum state pulses in the last 150 ns. As the different durations and amplitudes of the phase waveform patterns in a basic period, normal RF amplifiers are difficult to maintain without distortion after amplifying the modulation signal with a high gain which inevitably contributes to QBER in $X$-basis, especially the driving voltage of the phase modulator is approximate to 9 V peak-peak value, therefore by applying two linear amplifiers to drive the phase modulators, we can keep the error rate in $X$-basis less than 4\%.

Following the phase modulation, the two users modulate the light to 5 different intensities with three IMs. During the basic period of 1 $\mu$s, Alice and Bob randomly modulate the pulse intensity to $\mu_x$, $\mu_x\in \{\mu_z, \mu_1, \mu_2, 0 \}$ on 15 signal pulses with a 1 ns duration and 29 ns interval in the first 450 ns, $\mu_z$ is the intensity of the signal state, $\mu_1$ is the intensity of the weak decoy state, $\mu_2$ is the intensity of the strong decoy state, and $0$ is the vacuum state, in the next 400 ns, they modulate the intensity to $\mu_{ref}$ on 4 reference pulses with a each 100 ns duration, finaly they modulate the intensity to $0$ on vacuum state pulses in the last 150 ns as the recovery time for the superconducting nanowire single-photon detectors (SNSPDs).

Here, the signal pulses include the signal state pulse, weak decoy state pulse, strong decoy state pulse and vacuum state pulse. The first IM modulates light to 5 different intensities, the maximum intensity pulses are used as phase reference pulses, while the second as the signal state, the third as strong decoy state, the fourth as weak decoy state, and the minimum as vacuum state. The last two IMs modulates the intensities of the phase reference pulses and signal pulses to the designed ratio to ensure that the output signal intensities are agreement with the theoretical requirements and the reference detections are high enough for phase compensation; In order to reduce the detection probability of the noise, the signal pulse width is set to 1 ns of the last IM, additionally, the signal pulse width is set to 3 ns and 2 ns of the first two IMs to ensure the modulation signal pulses of the three IMs are well overlapped. The reference pulse width of the 3 IMs is set to 100 ns to ensure the maximum reference detections. All the IMs placed in a thick foam box to reduce the fluctuations of the ambient temperature in the lab. And all the IMs block the light when the pulse is a ``vacuum'' state, or is a ``not-sending'' state in the $Z$-basis. Before sent away from the secure zone of Alice and Bob, the signal intensities are attenuated to the single-photon level by an attenuator.

\section{Detailed Experimental Parameters}
Consistently, we set the length of the fiber between Alice and Charlie to be the same length between Bob and Charlie, the detailed parameters which includes the optical efficiencies of the fibers and optical devices at the measurement site, the proportions and intensities of each states for each fiber lengths are summarized in Tab.~\ref{Tab:Parameters}. The ``fiber length'' in the table is the total fiber length, where $\mu_1$, $\mu_2$ and $\mu_z$ are the intensities for the decoy states and the signal state. Here, we set the width of the signal pulse to 1 ns, while the effective width of the phase reference pulses to 100 ns, and the corresponding intensity of each 1 ns denoted by $\mu_{ref}$.

In the experiment over different fiber distances, based on the performance of modulation systems as well as detection noise, the ratio of sending and the total number ($N_{total}$) of signal pulses are different. The ratio of sending $X$ ($Z$) basis is $p_X$ ($p_Z$). In the $X$ basis, the ratio of sending vacuum ($0$), weak decoy state ($\mu_1$) and strong decoy state ($\mu_2$) are $p_0$, $p_1$ and $p_2$. In $Z$ basis, the fractions of ``sending'' and ``not-sending'' pulses are $p_{z1}$ and $p_{z0}$, respectively.

Finally, according to the above parameters, we calculated the intensity at Alice's (Bob's) output. Taking the optics  efficiencies and detection efficiencies into account, we also estimated the detection counts.

\begin{table*}[htb]
\centering
  \caption{Experimental parameters for the different fiber lengths.}
\begin{tabular}{c|cccc|ccc}
\hline
Fiber Length & 350 km & 408 km & 509 km \\
\hline
$\mu_1$  & 0.1 & 0.150 & 0.1 \\
$\mu_2$  & 0.411 & 0.393 & 0.384 \\
$\mu_z$  & 0.496 & 0.495 & 0.447 \\
$\mu_{ref}$ & 29.382 & 36.658 & 247.3 \\ 
\hline
$p_X$  & 0.134 & 0.113 & 0.224  \\
$p_Z$  & 0.866 & 0.887 & 0.776  \\
$p_0$  & 0.036 & 0.028 & 0.077  \\
$p_1$  & 0.898 & 0.874 & 0.850  \\
$p_2$  & 0.066 & 0.098 & 0.073  \\
$p_{z0}$ & 0.725 & 0.724 & 0.732 \\
$p_{z1}$ & 0.275 & 0.276 & 0.268 \\
\hline
Reference Width (ns)  & 100 & 100 & 100  \\
Output Intensity (nW)  & 1.503  & 1.875  & 12.65 \\ 
Detections (MHz)  & 1.85 & 1.85 & 1.85 \\
\hline
\end{tabular}
\label{Tab:Parameters}
\end{table*}

\section{The Re-Rayleigh scattering in fiber}
When the light wave with the wavelength of $\lambda_0$ travels through an optical fiber, most of the energy propagates forward, experiencing a loss from the fiber's attenuation, while a small portion of the light undergoes scattering, which is the dispersal of a light beam into a multitude of other beams radiated in a range of directions. As the reference pulses need to apply the same wavelength as the signal and share the same fiber in a time-multiplexing way in the experiment of SNS-TF-QKD, the scattering of strong reference pulses in fiber inevitably induce noise. There are several scattering mechanisms that operate in the optical fiber mainly includes elastic scattering and inelastic scattering.

\begin{figure}[tbh]
\centering
\resizebox{9cm}{!}{\includegraphics{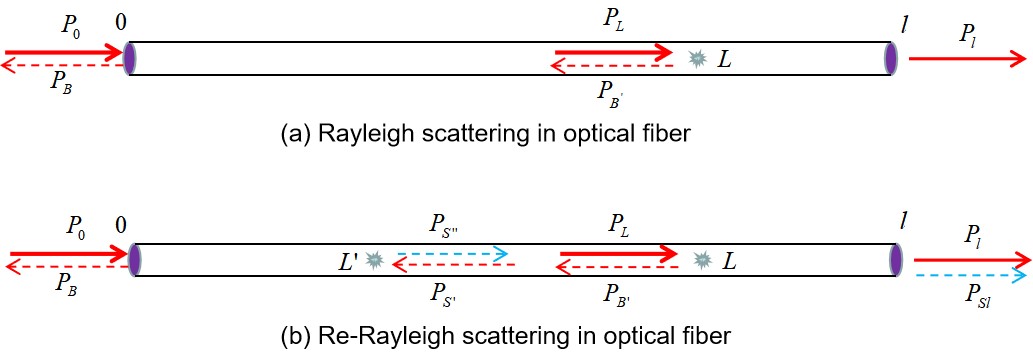}}
\caption{Rayleigh scattering in fiber.}
\label{Fig:rayleighscattering}
\end{figure}

In an optical fiber, the elastic scattering is mainly Rayleigh scattering, the wavelength of the Rayleigh scattering is same with the propagating beam. As shown in Fig.~\ref{Fig:rayleighscattering}(a), at the point 0, the seed light with intensity of $P_0$ is sent into the single mode fiber (SMF), and $P_0$ is attenuated to $P_L$ at the point $L$ due to the intrinsic loss of the fiber:
\begin{equation}
P_L=P_0e^{{-\alpha}L}
\end{equation}
where alpha is the intrinsic loss coefficient of fiber.

Then, at the point of $L$, affected by impurity defects in the SMF, the light with intensity of $P_L$ is backward scattered with Rayleigh scattering to the intensity of $P_{B'}$, further the $P_{B'}$ is attenuated to $P_B$ at the point of 0. Thus, the intensity of $dP_B$ at the point of 0 caused by Rayleigh scattering in the range of $dL$ at the point L can be described as:
\begin{equation}
dP_B=dP_{B'}e^{{-\alpha}L}=P_LSe^{{-\alpha}L}{dL}=P_0Se^{{-2\alpha}L}dL
\end{equation}
The total intensity of the backward Rayleigh scattering can be described as:
\begin{equation}
P_B=\int_{0}^{L} \\dP_B=\frac{P_0S}{2\alpha}(1-e^{{-2\alpha}L})
\end{equation}
where $S=\frac{2\alpha P_B}{P_0(1-e^{{-2\alpha}L})}$ is the coefficient of backward Rayleigh scattering. If $L\geq$50 km, $1-e^{{-2\alpha}L}\approx1$, $S=\frac{2\alpha P_B}{P_0}$.

Furthermore, as there are many impurity defects in optical fiber in the range of point 0 and point L, the backward scattered light at point $L$ with the intensity of $P_{B'}$ would be Rayleigh scattered again which is affected by the impurity defects at point $L'$. As shown in Fig.~\ref{Fig:rayleighscattering}(b), we define the Rayleigh scattering of the backward scattered light originating from the Rayleigh scattering of the seed light as Re-Rayleigh scattering in optical fiber. At the point of $L'$, the intensity $P_{S'}$ of the backward scattered light from point $L$ can be described as:
\begin{equation}
P_{S'}=P_{B'}e^{{-\alpha}(L-L')}
\end{equation}
and the corresponding intensity of the Re-Rayleigh scattering at the point of $L'$ can be expressed as:
\begin{equation}
P_{S''}=P_{S'}S=P_{B'}Se^{{-\alpha}(L-L')}
\end{equation}
Thus, the intensity of $dP_{Sl}$ at the point of $l$ caused by Re-Rayleigh scattering in the range of $dL$ at the point $L$ can be described as:
\begin{equation}
\begin{split}
\label{E1}
dP_{Sl}&=dP_{S''}e^{{-\alpha}(l-L')}=dP_{S'}Se^{{-\alpha}(l-{L'})}d{L'}\\
&=P_LS^2e^{{-\alpha}(L-{L'})}e^{{-\alpha}(l-{L'})}dLd{L'}\\
&=P_0S^2e^{{-\alpha}L}e^{{-\alpha}{(L-{L'})}}e^{{-\alpha}(l-{L'})}dLd{L'}
\end{split}
\end{equation}
As the polarization of the Re-Rayleigh scattering is random in optical fiber, half of the probability of which causes noise in measurement station, thus the total intensity of the Re-Rayleigh scattering noise $P_{srs}$ can be described as:
\begin{equation}
\begin{split}
\label{E1}
P_{srs}&=\frac{1}{2}P_{Sl}\\
&=\int_{0}^{l} \int_{0}^{L} \ \frac{1}{2}P_0S^2e^{{-\alpha}L}e^{{-\alpha}{(L-{L'})}}e^{{-\alpha}(l-{L'})}d{L'}\ dL\\
&=\frac{P_0S^2}{4\alpha}e^{-{\alpha}l}[l+\frac{e^{-2{\alpha}l}}{2\alpha}-\frac{1}{2\alpha}]
\end{split}
\end{equation}

\begin{figure}[tbh]
\centering
\resizebox{9cm}{!}{\includegraphics{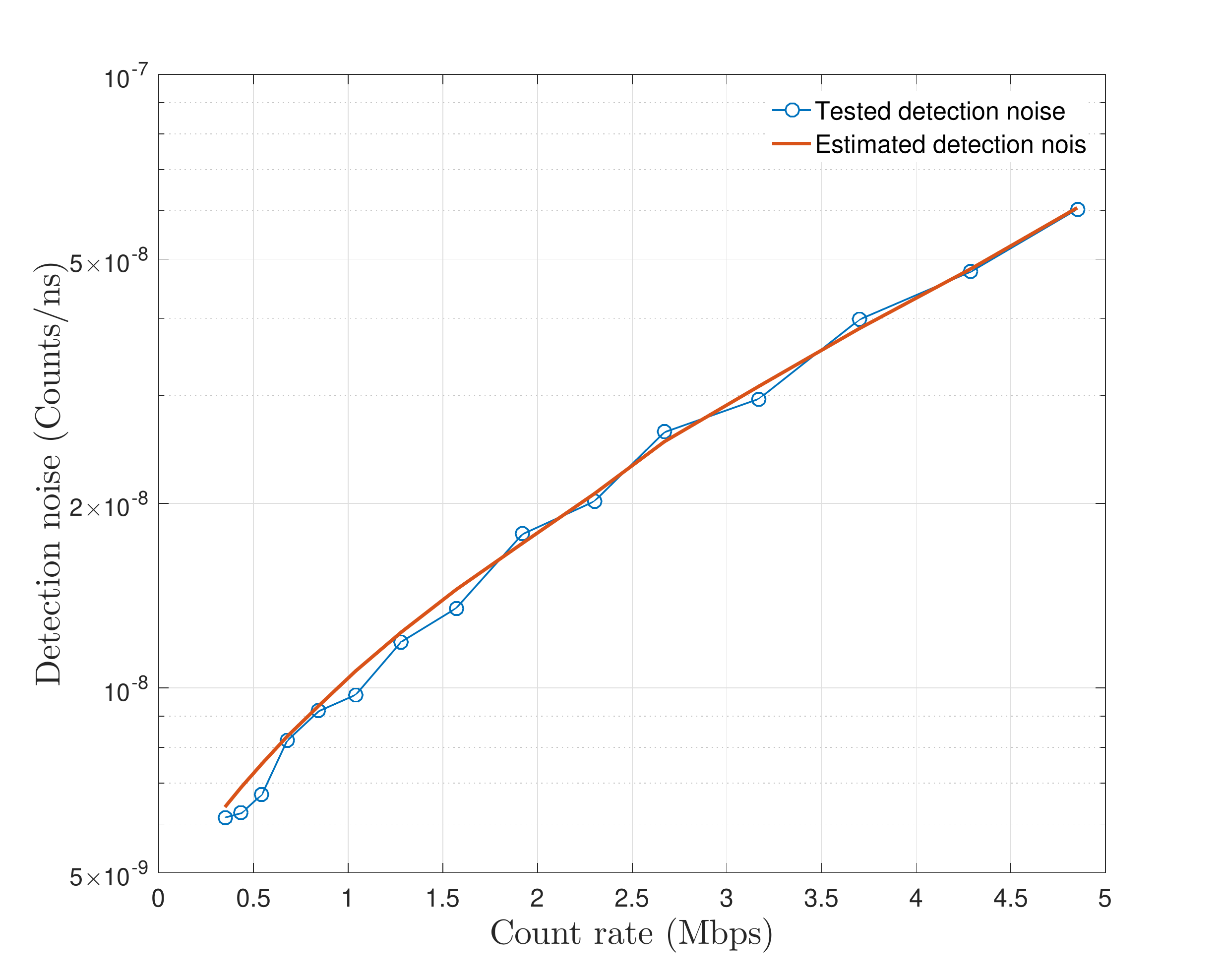}}
\caption{Detection noise caused by Re-Rayleigh scattering corresponding to the different count rates with 250km standard optical fiber. The blue circles show the experimental results and the orange curve shows the theoretical estimation results.}
\label{Fig:detectionnoise}
\end{figure}

Additionally, due to the intensity of the strong reference pulses is much less than 1 mw, the inelastic scattering consists of spontaneous Brillouin scattering and spontaneous Raman scattering in fiber. Here, the inelastic spontaneous scattering is very weak, we can only consider the forward inelastic scattering noise. Compared to Rayleigh scattering, the intensity of the forward spontaneous Brillouin scattering is two to three orders of magnitude lower with a frequency shift less than 1 GHz, while the forward spontaneous Raman scattering is three to five orders of magnitude lower with a THz frequency shift~\cite{kumar2014fiber}. The forward spontaneous Brillouin scattering noise from strong reference pulses is no effect to signal as the dispersion of the noise beam is much less than 14.5 ns (the minimum separation between signal pulses and reference pulses). Although the forward spontaneous Raman scattering noise would overlaps the signal pulses, it can be blocked by using a narrow band filter such as DWDM. Thus, only considering the inevitable noise in our experimental system, we can describe the detection noise as the following formula:
\begin{equation}
d=\frac{P_0S^2}{4{E_\nu}\alpha}e^{-{\alpha}l}[l+\frac{e^{-2{\alpha}l}}{2\alpha}-\frac{l}{2\alpha}]+D_c
\label{eq:detectionnoise}
\end{equation}
where the first term is described as the Re-Rayleigh scattering noise, and the second term is the dark count of SNSPD; $d$ is the detection noise count, $E_\nu$ is the photon energy, $l$ is the fiber length, $\alpha$ is the loss coefficient of fiber, $P_0$ is the intensity of light sent into the fiber, $D_c$ is the dark count of SNSPD. Ultimately, as shown in Fig.~\ref{Fig:detectionnoise}, we tested the detection noise corresponding to the different count rates with a single source over 250 km standard optical fiber, also, in the main text, by applying two remotely locked laser sources, we tested the detection noise over 408 km and 509 km ultra-low loss fiber with 2 MHz count rate in our experiment, which are almost agreement with the estimation results.

In a brief conclusion, in the experiment of SNS-TF-QKD, the scattering from the strong reference pulses in fiber prevents the detected noise from being as low as the dark count of the SNSPDs over long fiber links. Here, there is no need to consider the forward spontaneous Brillouin scattering noise as the dispersion of the noise beam without overlapping to the signal pulses; And the forward spontaneous Raman scattering noise can be blocked by using a narrow band filter. Though the Re-Rayleigh scattering noise is acceptable at 500 km scale, as an intrinsic noise, it deteriorates with distance if we want to maintain the reference detection rate for an accurate estimation of relative phase that inevitably degrades the performance of the TF QKD and will become a severe problem in further longer distance.

\section{Detailed Experimental Results}
In Tabs.~\ref{Tab:Characterization}, we characterized our experimental system which includes the efficiency of the SNSPDs (SNSPD-ch1/SNSPD-ch2), the insertion loss of the signal fiber links ($Loss_{FiberA}$, $Loss_{FiberB}$) the transmittance of the optical elements consist of the polarization controller (PC-A/PC-B), the circulator (CIR-A/CIR-B), the dense wavelength division multiplexing (DWDM-A/DWDM-B), polarization beam splitter (PBS-A/PBS-B) and the polarization maintaining beam splitter (BS-A-ch1/BS-A-ch2 and BS-B-ch1/BS-B-ch2) in the measurement station.

\begin{table*}[htb]
\centering
  \caption{Experimental parameters for different fiber lengths.}
\begin{tabular}{c|ccc}
\hline
Fiber Length    & 350 km  & 408 km  & 509 km  \\
$Loss_{FiberA}$	& 33.00dB & 34.38dB & 42.50dB     \\
$Loss_{FiberB}$	& 33.20dB & 34.04dB & 42.10dB   \\
\hline
PC-A & \multicolumn{3}{c}{94.2\%}\\
PC-B & \multicolumn{3}{c}{92.8\%}\\
\hline
CIR-A & \multicolumn{3}{c}{84.7\%}\\
CIR-B & \multicolumn{3}{c}{85.2\%}\\
\hline
DWDM-A & \multicolumn{3}{c}{89.3\%}\\
DWDM-B & \multicolumn{3}{c}{88.7\%}\\
\hline
PBS-A & \multicolumn{3}{c}{91.1\%}\\
PBS-B & \multicolumn{3}{c}{86.5\%}\\
\hline
BS-A-ch1 & \multicolumn{3}{c}{36.9\%}\\
BS-A-ch2 & \multicolumn{3}{c}{38.6\%}\\
BS-B-ch1 & \multicolumn{3}{c}{39.1\%}\\
BS-B-ch2 & \multicolumn{3}{c}{41.4 \%}\\
\hline
SNSPD-ch1 & \multicolumn{3}{c}{58.0\%}\\
SNSPD-ch2 & \multicolumn{3}{c}{56.0\%}\\
\hline
\end{tabular}
\label{Tab:Characterization}
\end{table*}

In order to obtain a sufficiently good performance of the interference with a higher visibility, we set a detection window which is narrower than the pulse width to the signal pulses. At the same time, in Alice's (Bob's) station, limited by the AWG, the time jitter of the signal pulse is about 100 ps, further the jitter of the signal pulses deteriorates along with the transmission fiber links. Caused by jitter of the signal pulses, non-overlap between the signal pluses and detection windows added the insertion loss of Charlie's measurement devices. The fraction of the data falling within the detection window is represented by the parameter $r_{gate}$. Note that all the detections are filtered according to the digital window and the estimation success probability $rc$ by Charlie, before the detections are announced. Here, the fraction of the data falling within the detection window of the experiment of 350 km, 408 km, and 509 km are about 51\%, 48\% and 35\% respectively.

For each fiber length, the experimental results are summarized in Tab.~\ref{Tab:Result}. In the table, we denote $n_1$(Before AOPP) ($n_1$(After AOPP)) as the  number of the untagged bits in $Z_1$-windows, $e_1^{ph}$(Before AOPP) ($e_1^{ph}$(After AOPP)) as the phase-flip error rate, and QBER$(Z-{Before})$ (QBER$(Z-{After})$) as the bit-flip error rate before (after) the bit error rejection by active odd parity pairing (AOPP).
With the best accepted phase difference $Ds$ (in degrees) and the total number of signal pulses $N_{total}$, we calculate the error rates in the $Z$ bases are given by ``QBER$(Z-{Before})$'', ``$e_1^{ph}$(Before AOPP)'', and the error rates in the $X$ bases are given by ``QBER(X11)'' for the weak decoy states ``11'' before AOPP. Similarly, applying the bit error rejection by AOPP, the error rates in the $Z$ bases are given by ``QBER$(Z-{After})$'' and ``$e_1^{ph}$(After AOPP)''. Then, based on the accepted phase estimation success probabilities $rc$, we optimized  the different phase difference ranges $Ds$ to maximize the final key rate $R$.

Furthermore, in the following rows, we list the numbers of pulses Alice and Bob sent in different decoy states, labelled as ``Sent-ABCD'', where ``A'' (``B'') is ``X'' or ``Z'' indicating the basis which Alice (Bob) has used; ``C'' (``D'') is ``0'', ``1'',  ``2'' or  ``3'', indicating the intensity which Alice (Bob) has chosen within ``vacuum'', ``$\mu_1$'', ``$\mu_2$'' or ``$\mu_z$''.

The total number sent pulses of Alice and Bob is listed as ``Sent-AB''. As with the numbers of sent pulses, the numbers of detections are listed as ``Detected-ABCD''. The total valid detections reported by Charlie is denoted as ``Detected-Valid-ch'', where ``ch'' can be ``Det1'' or ``Det2'' indicating the responsive detector of the recorded counts. The table also gives the numbers of detections falling within the accepted difference range $Ds$, listed as ``Detected-ABCD-Ds-Ch'', where ``Ds'' indicates that only the data within the accepted range $Ds$ are counted, ``Ch'' indicates the detection channel. The numbers of correct detections are listed as ``Correct-ABCD-Ds-Ch'', and used to calculate the $X$ basis error rate. The optimized acceptance ranges are listed on the top lines of this table.

\begin{table*}[htb]
\centering
  \caption{Experimental results for different fiber lengths.}
\begin{tabular}{c|ccc}
\hline
Fiber Length & 350 km & 408 km & 509 km \\
\hline
$N_{total}$	 & $3.05\times 10^{11}$ & $5.20\times 10^{11}$ & $1.55\times 10^{12}$\\
$R$			 & $6.34\times 10^{-7}$ & $3.22\times 10^{-7}$ & $1.79\times 10^{-8}$ \\
\hline
$n_1$(Before AOPP)		 & 2444570 & 2828500 & 729886 \\
$n_1$(Atter AOPP)		 & 412655 & 411515 & 128059 \\
$e_1^{ph}$(Before AOPP)  & 5.82\% & 6.78\% & 6.46\%  \\
$e_1^{ph}$(After AOPP)   & 11.13\% & 12.83\% & 12.40\%  \\
\hline
QBER$(Z-{Before})$ 	 & 27.33\% & 27.47\% & 27.07\%  \\
QBER$(Z-{After})$ 	 & 0.072\% & 0.076\% & 0.922\%  \\
\hline
QBER(X11)	 & 3.3\% & 3.0\% & 3.8\%  \\
QBER(X22)	 & 1.8\% & 3.0\% & 2.2\%  \\
\hline
$rc$		 & 1 			& 1 			& 0.5	\\
$Ds$		 & $15^\circ$		& $12^\circ$ 	& $15^\circ$ \\
$r_{rc}$	 & 1			& 1			& 0.9996 \\
$r_{gate}$	 & 0.51 			& 0.48			& 0.35\\
\hline
Sent-ZZ		 & 228395800000	& 409130800000	& 935934600000		\\
Sent-ZX00	 & 912200000	& 1079200000	& 15419600000		\\
Sent-ZX01	 & 23068800000	& 33283400000	& 168177400000		\\
Sent-ZX02	 & 1719800000	& 3724600000	& 14383000000		\\
Sent-ZX30	 & 345200000	& 399800000		& 5596200000				\\
Sent-XZ00	 & 929000000	& 1074000000	& 15330800000		\\
Sent-XZ10	 & 23040800000	& 33200000000	& 168255400000	\\
Sent-XZ20	 & 1690000000	& 3752600000	& 14364400000		\\
Sent-XZ03	 & 352800000	& 408600000		& 5571400000				\\
Sent-XX00	 & 7800000		& 5200000		& 466600000		\\
Sent-XX01	 & 178200000	& 154000000		& 5163400000 \\
Sent-XX02	 & 17200000		& 17000000		& 438000000 \\
Sent-XX10	 & 182200000	& 163800000		& 5111800000 \\
Sent-XX20	 & 13200000		& 19600000		& 442200000 \\
Sent-XX11	 & 4409800000	& 5074800000	& 56133000000 	\\
Sent-XX22	 & 22200000		& 62400000		& 412400000 \\
\hline
Detected-Valid-Det1	& 3840607		& 5073367		& 1407071	 \\	
Detected-Valid-Det2	& 3706688		& 4954148		& 1353476 \\
\hline
Detected-ZX00	& 22		& 15		& 239	  \\
Detected-ZX01	& 214060	& 324071	& 120954  \\
Detected-ZX02	& 62725 	& 109410	& 36537	  \\
Detected-ZX30	& 16168 	& 14097 	& 20011  \\
Detected-XZ00	& 14		& 17		& 206	  \\
Detected-XZ10	& 215668	& 346605	& 132391	  \\
Detected-XZ20	& 61585 	& 100798	& 41210	  \\
Detected-XZ03	& 15631 	& 13294 	& 17545	 	 \\
Detected-XX00	& 0 		& 0		    & 6		 		 \\
Detected-XX01	& 1688  	& 1441  	& 3761	 	 \\
Detected-XX02	& 681   	& 466	    & 1139	 	 \\
Detected-XX10	& 1736  	& 1656  	& 3980	 	 \\
Detected-XX20	& 495   	& 571   	& 1324	 	 \\
Detected-XX11	& 82448 	& 102201	& 84034	 	 \\
Detected-XX22	& 1575  	& 3606  	& 2198	 	 \\
\hline
Detected-XX11-Ds-Ch1	& 7338	& 7341	& 7341  \\
Detected-XX11-Ds-Ch2	& 7071	& 7107	& 7059	 \\
Correct-XX11-Ds-Ch1		& 7110	& 7139	& 7105  \\
Correct-XX11-Ds-Ch2		& 6841	& 6880	& 6750  \\
\hline
Detected-ZZError	& 1577237	& 2107983	& 458709	   \\
Detected-ZZCorrect	& 4194347	& 5565128	& 1235589	  \\
\hline
\end{tabular}
\label{Tab:Result}
\end{table*}

Additionally, we calculate the QBERs of the $X$-basis over 509 km when Alice and Bob sent decoy states of $\mu_1$ and $\mu_2$ with the different detection counts according to the different phase difference range ($Ds$) and the accepted phase estimation success probabilities $rc$, which are listed in Tabs.~\ref{Tab:ResultQBERXX11}, Tabs.\ref{Tab:ResultQBERXX22}, Tabs.~\ref{Tab:ResultDetXX11} and \ref{Tab:ResultDetXX22} respectively. Finally, as shown in Tab.~\ref{Tab:ResultKeyRate}, by searching though ranges of parameter values to find the optimized key rate, we extract the secure key rates with different parameter values.

\begin{table*}[htb]
\centering
  \caption{$X$ basis QBERs for decoy state $\mu_1$ with 509 km ultra-low loss fiber.}
\begin{tabular}{c|cccccccc}
\hline
RC$\mid$Ds/2   & deg=2$^\circ$	& deg=5$^\circ$	& deg=8$^\circ$	& deg=10$^\circ$	& deg=12$^\circ$	& deg=15$^\circ$	& deg=30$^\circ$\\
\hline
rc=0.01	& 3.0\% & 3.1\% & 3.2\% & 3.3\% & 3.6\% & 3.6\% & 6.1\%  \\	
rc=0.05	& 3.0\% & 3.2\% & 3.2\% & 3.5\% & 3.7\% & 3.8\% & 6.1\%  \\
rc=0.10	& 2.9\% & 3.2\% & 3.4\% & 3.5\% & 3.7\% & 3.7\% & 6.1\%  \\	
rc=0.50	& 2.9\% & 3.2\% & 3.4\% & 3.5\% & 3.7\% & 3.8\% & 6.2\%  \\
rc=1.00	& 2.9\% & 3.3\% & 3.4\% & 3.5\% & 3.7\% & 3.8\% & 6.2\%  \\
\hline
\end{tabular}
\label{Tab:ResultQBERXX11}
\end{table*}

\begin{table*}[htb]
\centering
  \caption{$X$ basis QBERs for decoy state $\mu_2$ with 509 km ultra-low loss fiber.}
\begin{tabular}{ccccccccc}
\hline
RC$\mid$Ds/2 & deg=2$^\circ$	& deg=5$^\circ$	& deg=8$^\circ$	& deg=10$^\circ$	& deg=12$^\circ$	& deg=15$^\circ$	& deg=30$^\circ$\\
\hline
rc=0.01	& 0 & 0     & 1.1\% & 1.9\% & 1.5\% & 2.5\% & 4.0\% \\
rc=0.05	& 0 & 1.0\% & 1.2\% & 2.1\% & 1.7\% & 2.1\% & 4.6\% \\
rc=0.10	& 0 & 0.9\% & 1.1\% & 1.9\% & 1.5\% & 1.9\% & 4.7\% \\
rc=0.50	& 0 & 1.6\% & 1.6\% & 2.3\% & 1.9\% & 2.2\% & 4.9\% \\
rc=1.00	& 0 & 1.6\% & 1.6\% & 2.3\% & 1.9\% & 2.2\% & 4.9\% \\
\hline
\end{tabular}
\label{Tab:ResultQBERXX22}
\end{table*}

\begin{table*}[htb]
\centering
  \caption{$X$ basis Detections for decoy state $\mu_1$ with 509 km ultra-low loss fiber.}
\begin{tabular}{c|cccccccc}
\hline
RC$\mid$Ds/2 & deg=2$^\circ$  & deg=5$^\circ$	& deg=8$^\circ$	& deg=10$^\circ$	& deg=12$^\circ$	& deg=15$^\circ$	& deg=30$^\circ$\\
\hline
rc=0.01	& 1078 & 2436  & 3827  & 4736 & 5667  & 6945  & 13806 \\
rc=0.05	& 2016 & 4496  & 7027  & 8645 & 10297 & 12693 & 25040 \\
rc=0.10	& 2231 & 4963  & 7748  & 9515 & 11350 & 13955 & 27477 \\
rc=0.50 & 2286 & 5114  & 7990  & 9814 & 11710 & 14400 & 28354 \\
rc=1.00 & 2287 & 5117  & 7994  & 9818 & 11715 & 14405 & 28365 \\
\hline
\end{tabular}
\label{Tab:ResultDetXX11}
\end{table*}

\begin{table*}[htb]
\centering
  \caption{$X$ basis Detections for decoy state $\mu_2$ with 509 km ultra-low loss fiber.}
\begin{tabular}{c|cccccccc}
\hline
RC$\mid$Ds/2 & deg=2$^\circ$	& deg=5$^\circ$	& deg=8$^\circ$	& deg=10$^\circ$	& deg=12$^\circ$	& deg=15$^\circ$	& deg=30$^\circ$	\\
\hline
rc=0.01 & 27 & 52  & 92  & 108 & 131 & 162 & 351 \\
rc=0.05 & 57 & 105 & 166 & 195 & 236 & 285 & 608 \\	
rc=0.10 & 62 & 115 & 183 & 214 & 259 & 313 & 675 \\	
rc=0.50 & 64 & 122 & 191 & 222 & 267 & 324 & 698 \\	
rc=1.00 & 64 & 122 & 191 & 222 & 267 & 324 & 698 \\	
\hline
\end{tabular}
\label{Tab:ResultDetXX22}
\end{table*}

\begin{table*}[htb]
\centering
  \caption{Key Rate for different parameters with 509 km ultra-low loss fiber.}
\begin{tabular}{c|cccccccc}
\hline
RC$\mid$Ds/2 & deg=2$^\circ$	& deg=5$^\circ$	& deg=8$^\circ$	& deg=10$^\circ$	& deg=12$^\circ$	& deg=15$^\circ$	& deg=30$^\circ$\\
\hline
rc=0.01	& $0$ & $3.29\times 10^{-9}$ & $4.95\times 10^{-9}$ & $5.35\times 10^{-9}$ & $5.05\times 10^{-9}$ & $5.64\times 10^{-9}$ & $1.35\times 10^{-9}$ \\
rc=0.05	& $4.02\times 10^{-9}$ & $1.16\times 10^{-8}$ & $1.37\times 10^{-8}$ & $1.45\times 10^{-8}$ & $1.42\times 10^{-8}$ & $1.48\times 10^{-8}$ & $6.28\times 10^{-9}$ \\
rc=0.10	& $6.37\times 10^{-9}$ & $1.41\times 10^{-8}$ & $1.59\times 10^{-8}$ & $1.68\times 10^{-8}$ & $1.64\times 10^{-8}$ & $1.73\times 10^{-8}$ & $7.59\times 10^{-9}$  \\
rc=0.50	& $7.38\times 10^{-9}$ & $1.47\times 10^{-8}$ & $1.67\times 10^{-8}$ & $1.76\times 10^{-8}$ & $1.72\times 10^{-8}$ & $1.79\times 10^{-8}$ & $7.71\times 10^{-9}$  \\
rc=1.00 & $7.40\times 10^{-9}$ & $1.44\times 10^{-8}$ & $1.65\times 10^{-8}$ & $1.75\times 10^{-8}$ & $1.70\times 10^{-8}$ & $1.78\times 10^{-8}$ & $7.63\times 10^{-9}$  \\
\hline
\end{tabular}
\label{Tab:ResultKeyRate}
\end{table*}

\end{document}